\documentclass[a4paper,fleqn]{cas-dc}

\usepackage[numbers]{natbib}
\usepackage{geometry}
\usepackage{graphics}

\usepackage{ragged2e}
\usepackage{caption}
\usepackage{subcaption}
\usepackage{threeparttable}
\usepackage{cleveref}
\usepackage{enumitem}
\usepackage{gensymb} 

\usepackage{tabularx}
\usepackage{notoccite}
\usepackage{balance}

\begin{document}
\let\WriteBookmarks\relax
\def\floatpagepagefraction{1}
\def\textpagefraction{.001}
\shorttitle{Centralized MPC for Thermal Comfort and Residential Energy Management}
\shortauthors{Seal, S. et~al.}

\title [mode = title]{Centralized Model Predictive Control Strategy for Thermal Comfort and Residential Energy Management}                      
%
%

\author[1]{Sayani Seal}[%
    orcid=0000-0002-0982-8272]
\cormark[1]
%
\credit{Conceptualization of this study, Formal analysis, Investigation, Methodology, Software,  Writing - original draft preperation}
%

\author[1]{Benoit Boulet}

\credit{Conceptualization of this study, Supervision, Visualization, Writing - review and editing}

\author[2]{Vahid R. Dehkordi}
%
\credit{Supervision, Writing - review and editing}
\address[1]{McGill University, Montreal, Quebec, Canada}
\address[2]{NRCan-CanmetENERGY, Varennes, Quebec, Canada}

%

\cortext[cor1]{Corresponding author}
%

\begin{abstract}
A novel centralized model predictive control (MPC) is proposed for comfort and energy management in residential building. The residential setup used here is equipped with a photovoltaic (PV) solar system and a stationary home battery unit. An air-to-air multi-split heat pump (HP) is used as the primary heating system. The electric baseboard (BB) unit in each zone is used as a secondary system. The MPC is simultaneously responsible for controlling the heating inputs of the HP and BB units for comfort management, as well as for the control of energy flow between the PV, the home battery and the bidirectional grid system. Variable Time-of-Use (ToU) rates are considered for the energy cost calculation and Feed-in-Tariff (FiT) is considered for selling energy to the grid. A 13.5\% reduction in the energy cost is achieved with the centralized MPC as compared to a rule based energy management strategy. The solar energy generation and battery storage contribute to approximately 31\% saving.
\end{abstract}



\begin{keywords}
Centralized model predictive control (MPC) \sep Home energy management \sep Solar panels \sep Home battery
\end{keywords}

\maketitle


\section{Introduction}

The heating system is an integral part of the heating, ventilation and air-conditioning (HVAC) system that constitutes the major part of the residential electrical load. But the electrical loads, such as, daily hot water requirement, electrical appliances, lighting, also add up to the net residential electricity consumption. In this article, an inclusive study of the overall energy management of a residential building is presented considering renewable energy generation and storage. A centralized model predictive control (MPC) strategy is proposed which simultaneously handles the comfort management along with the optimization of energy flow among the different components of the grid-connected residential energy network.

The benefits of a grid connected PV generation system coupled with a battery storage unit is studied in \cite{SANIHASSAN2017422_EMS} based on different electricity rates, feed-in tariff (FiT) incentive and battery storage unit cost. In this paper, an existing residential PV system is optimized with and without battery storage to maximize FiT revenue considering three import tariffs, i.e.,  a flat rate, a Time-of-Use rate based and a wholesale electricity tariff in the first part of the work. Subsequently, a sensitivity analysis is performed to quantify the impact of the battery storage unit cost on the objective function and the results are used to determine the optimal battery capacity. Finally, a cost analysis is presented by considering the unit cost of battery storage and a new PV module in maximizing the FiT revenue.

In \cite{Chen6575202_EMS} a task-based appliance scheduling scheme for a smart grid connected residential building is proposed considering both real-time and forecast information, time varying electricity price and weather data. Occupant comfort is also integrated in the same optimization. A finite-horizon MPC control is used as a building energy management controller (BEMC) for optimal scheduling of thermostatically controlled heating-cooling units and the non-thermal appliances based on their delay and power consumption flexibilities. Plug-in electric vehicle (PEV) batteries are also considered as a non-thermal appliance. For the non-thermal units, the MPC minimizes the electricity expense by controlling the ON/OFF status of the appliances and thereby optimizing the delay between the occupants' requests of using an appliance and its availability based on variable electricity rates. Performance constraints of the individual appliance are also considered to deliver quality service. For the thermal appliance scheduling, predictive mean vote (PMV) index is used as an indicator of occupants' comfort. An acceptable comfort range is defined as a function of the air temperature and the mean radiant temperature. The MPC schedules the thermal appliances, i.e., electric heater and air-conditioning systems to minimize the power consumption while considering the allowable PMV range as an optimization constraint.

However, in this paper, instead of considering the PMV index based comfort criterion as a constraint to the optimization \cite{Chen6575202_EMS}, the operative temperatures of the different thermal zones are continuously monitored to be maintained within an acceptable comfort band. For the energy management the appliance schedules are fixed and the MPC is bound to supply the load demand when needed. Instead of the load shifting as presented in \cite{Chen6575202_EMS}, here the MPC relies on the charging-discharging of the battery and the timely utilization of the renewable and grid based energy to reduce the energy cost.

\vspace*{0.1in}
\noindent The organization of this article is as follows. In \Cref{C5_sec_Objective} the objectives are presented. A detailed description of the case study house along with its heating systems and energy network arrangements are given in \Cref{C5_sec_Bungalow}. Subsequently, in \Cref{C5_Sys_ID_Val} the system identification and validation of the control-oriented house model are elaborated. The uncertainties in the disturbance inputs used in this simulation study and the optimization problem formulation are presented in \Cref{C5_sec_uncertainties} and \Cref{C5_sec_Prob}, respectively. Two case study scenarios are described in \Cref{C5_sec_PV_objectiveFunction_caseStudy} along with the optimization results presented in \Cref{C5_sec_result}. Finally, \Cref{C5_sec_conclusion} summarizes the work presented in this article. 
 

\section{Objectives}
\label{C5_sec_Objective}

\noindent This article proposes a novel inclusive MPC strategy that simultaneously handles different components of the building energy management system (BEMS) to achieve a predefined comfort while minimizing the energy cost. Scheduled daily electrical loads are considered to represent a rough estimate of the daily domestic energy consumption, considering the heating system as the primary load. The BEMS, discussed here, includes an on-site solar energy generation system and storage facility using a battery unit. A bidirectional grid connection is considered. The house heating system uses a combination of a multi-split heat pump with indoor fan units and supplementary electric baseboard units. The BEMS is centrally controlled by the MPC which performs a multi-variable optimization, with a single objective function, to efficiently manage the renewable energy usage, charging and discharging of the battery and load distribution between the energy efficient heat pumps and the baseboards to maintain the indoor operative temperature within given comfort bounds with a reduced heating energy consumption. The proposed approach contributes in combining two popular sects of building control engineering, namely the HVAC control for comfort management with low operative cost and energy management for building electricity consumption. 

A TRNSYS-MATLAB co-simulation is used to simulate the comfort and energy management case studies in this article. The MPC is designed in MATLAB and the optimized control signals are fed to a TRNSYS house model. Here, the TRNSYS house is treated as an alternative to the actual physical house and the MPC uses a simplified linear model of the TRNSYS house for temperature predictions. A similar co-simulation arrangement is described in detail in \cite{Seal_2019,Seal_thesis}.


\section{The case study house, its heating systems and the home energy network}
\label{C5_sec_Bungalow}

\begin{figure}[pos=h!]
\centering
    \includegraphics[scale = 0.32]{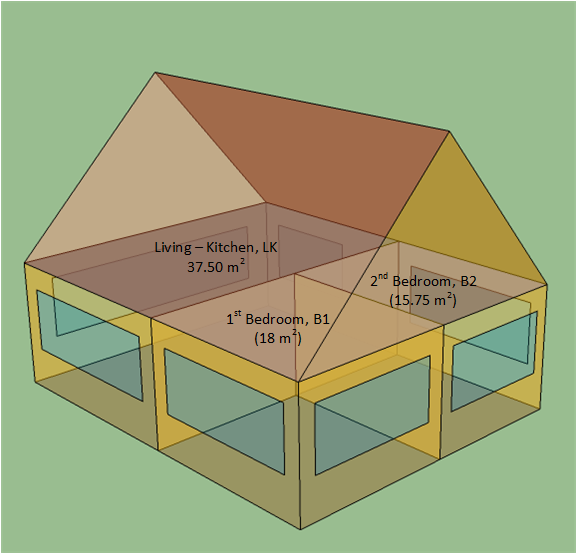}
 \captionof{figure}{Schematic diagram of the house with designated zones}
    \label{fig_C5_Bungalow}
\end{figure}

\noindent A single-story residence is designed for this simulation study using Google SketchUp and TRNSYS. The basic layout of the house is created in SketchUp, followed by modifications in TRNSYS to incorporate different physical attributes. 

\begin{table}[pos=h!]
    \caption{House envelope and RFH construction specifications}
    \label{C5_tab_Bungalow_Specs}
   {\setlength\extrarowheight{0.1pt}
    {\scriptsize\addtolength{\tabcolsep}{-6pt}
    \vspace*{0.1in}
    \begin{tabular*}{\tblwidth}{@{} CCC@{} } 
     \multicolumn{3}{c}{\ House Layout \ } \\ \toprule
     \multicolumn{1}{ l }{\textbf{\ Parameters \ }} &
     \multicolumn{1}{ l }{\textbf{\ Zones \ }} &
     \multicolumn{1}{ c }{\textbf{\ Specifications \ }} \\ \midrule
     \multicolumn{1}{ l }{\  Floor area \ } &
     \multicolumn{1}{ l }{\textbf{\ B1 \ }} &
     \multicolumn{1}{ c }{\ 18.00 m$^2$ \ } \\     
     \multicolumn{1}{ l }{ } &
     \multicolumn{1}{ l }{\textbf{\ B2 \ }} &
     \multicolumn{1}{ c }{\ 15.75 m$^2$ \ } \\  
     \multicolumn{1}{ l }{ } &
     \multicolumn{1}{ l }{\textbf{\ LK \ }} &
     \multicolumn{1}{ c }{\ 37.50 m$^2$ \ } \\  
     \multicolumn{1}{ l }{\ Height \ } &
     \multicolumn{1}{ l }{\textbf{\ All \ }} &
     \multicolumn{1}{ c }{\ 3.048 m \ } \\  
     \multicolumn{1}{ l }{\  Window area \ } &
     \multicolumn{1}{ l }{\textbf{\ B1-S \ }} &
     \multicolumn{1}{ c }{\ 5.49 m$^2$ \ } \\     
     \multicolumn{1}{ l }{ } &
     \multicolumn{1}{ l }{\textbf{\ B1-E \ }} &
     \multicolumn{1}{ c }{\ 4.80 m$^2$ \ } \\   
     \multicolumn{1}{ l }{ } &
     \multicolumn{1}{ l }{\textbf{\ B2-E \ }} &
     \multicolumn{1}{ c }{\ 4.29 m$^2$ \ } \\     
     \multicolumn{1}{ l }{ } &
     \multicolumn{1}{ l }{\textbf{\ B2-N \ }} &
     \multicolumn{1}{ c }{\ 5.49 m$^2$ \ } \\   
     \multicolumn{1}{ l }{ } &
     \multicolumn{1}{ l }{\textbf{\ LK-N \ }} &
     \multicolumn{1}{ c }{\ 3.75 m$^2$ \ } \\     
     \multicolumn{1}{ l }{ } &
     \multicolumn{1}{ l }{\textbf{\ LK-W \ }} &
     \multicolumn{1}{ c }{\ 9.15 m$^2$ \ } \\   
     \multicolumn{1}{ l }{ } &
     \multicolumn{1}{ l }{\textbf{\ LK-S \ }} &
     \multicolumn{1}{ c }{\ 6.10 m$^2$ \ } \\   \bottomrule
     \multicolumn{3}{c}{\ House Envelope \ } \\ \toprule
     \multicolumn{1}{ l }{\textbf{\ Parameters \ }} &
     \multicolumn{1}{ p{1cm} }{} &
     \multicolumn{1}{ c }{\textbf{\ Specifications \ }} \\ \midrule
     \multicolumn{1}{ l }{\ Roof RSI$^\ast$\tnote{$\ast$} \ } &
     \multicolumn{1}{ p{1cm} }{} &
     \multicolumn{1}{ l }{\ 6.897 m$^2$K/W (R39) \ } \\     
     \multicolumn{1}{ l }{\ External Wall RSI \ } &
     \multicolumn{1}{ p{1cm} }{} &
     \multicolumn{1}{ l }{\ 3.003 m$^2$K/W (R17) \ } \\      
     \multicolumn{1}{ l }{\ Internal Wall RSI \ } &
     \multicolumn{1}{ p{1cm} }{} &
     \multicolumn{1}{ l }{\ 0.513 m$^2$K/W (R3) \ } \\       
     \multicolumn{1}{ l }{\ Slab RSI \ } &
     \multicolumn{1}{ p{1cm} }{} &
     \multicolumn{1}{ l }{\ 0.319 m$^2$K/W (R2) \ } \\      
     \multicolumn{1}{ l }{\ Window U-value \ } &
     \multicolumn{1}{ p{1cm} }{} &
     \multicolumn{1}{ l }{\ 1.570 W/m$^2$K \ } \\       
     \multicolumn{1}{ l }{\ Infiltration \ } &
     \multicolumn{1}{ p{1cm} }{} &
     \multicolumn{1}{ l }{\ 2.5 $ACH_{50}$ \ } \\ \bottomrule 
     \end{tabular*}} \\ \vspace*{0.05in}
            \scriptsize{\hspace*{-1.5in}
            $^\ast$RSI : Thermal resistance in SI units}}
    \end{table}

The house has two bedrooms and a living-dining space with an open kitchen. The total floor area is 71.25 m$^2$, having a 37.50 m$^2$ of living-kitchen (LK) area and respectively 18 m$^2$ and 15.75 m$^2$ in the primary (B1) and secondary (B2) bedrooms. It has an attic with the South facing roof inclined at an angle of $45 \degree$. The PV panels are installed on this South facing roof. The attic is not considered as conditioned thermal space. The construction specifications are presented in \Cref{C5_tab_Bungalow_Specs}. 

\begin{figure}[pos=h!]
{    \centering
    \includegraphics[scale = 0.4]{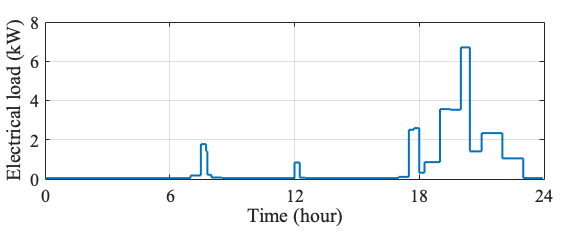}
 \captionof{figure}{Daily electrical load}
    \label{C5_fig_Bungalow_electrical_load}}
\end{figure}
    
The rated power draws for the electrical appliances are shown in \Cref{C5_tab_Bungalow_appliance_rating}. The percentages of convective heat gains \cite{App_heat_gain} from different appliances, contributing to the net internal gain, are also indicated in the Table. \Cref{C5_fig_Bungalow_electrical_load} represents a rough estimation of the total daily electrical energy consumption used in the simulation. It includes the lighting, scheduled loads of the electrical appliances, the energy requirement for daily hot water usage, the heat recovery ventilator unit and the PV regulator - inverter unit.

\begin{table}[pos=h!]
    \caption{Power ratings for electrical appliances}
    \label{C5_tab_Bungalow_appliance_rating} 
    {\setlength\extrarowheight{0.1pt}
    {\scriptsize\addtolength{\tabcolsep}{-6pt}
    \begin{tabular*}{\tblwidth}{@{} CCCCCC@{} } 
    \toprule 
     \multicolumn{1}{ l }{\textbf{\ Appliances \ }} &
     \multicolumn{1}{ p{1cm} }{} &
     \multicolumn{1}{ c }{\textbf{\ Rated Power (W) \ }} &
     \multicolumn{1}{ p{0.5cm} }{} &
     \multicolumn{1}{ c }{\textbf{\ Heat gain (\%) \ }} \\ \midrule 
     \multicolumn{1}{ l }{\ Clothes washer \ } &
     \multicolumn{1}{ p{0.5cm} }{} & 
     \multicolumn{1}{ l }{\ 1780 \ } & 
     \multicolumn{1}{ p{0.5cm} }{} &     
     \multicolumn{1}{ c }{\ 85 \ } \\     
     \multicolumn{1}{ l }{\ Dryer \ } &
     \multicolumn{1}{ p{0.5cm} }{} &
     \multicolumn{1}{ l }{\ 5300 \ } &
     \multicolumn{1}{ p{0.5cm} }{} &
     \multicolumn{1}{ c }{\ 15 \ } \\      
     \multicolumn{1}{ l }{\ Dishwasher  \ } &
     \multicolumn{1}{ p{0.5cm} }{} &
     \multicolumn{1}{ l }{\ 890 \ } &
     \multicolumn{1}{ p{0.5cm} }{} &
     \multicolumn{1}{ c }{\ 65 \ } \\    
     \multicolumn{1}{ l }{\ Refrigerator  \ } &
     \multicolumn{1}{ p{0.5cm} }{} &
     \multicolumn{1}{ l }{\ 720 \ } &
     \multicolumn{1}{ p{0.5cm} }{} &
     \multicolumn{1}{ c }{\ 100 \ } \\    
     \multicolumn{1}{ l }{\ Oven  \ } &
     \multicolumn{1}{ p{0.5cm} }{} &
     \multicolumn{1}{ l }{\ 2200 \ } &
     \multicolumn{1}{ p{0.5cm} }{} &
     \multicolumn{1}{ c }{\ 50 \ } \\  \bottomrule 
     \end{tabular*}}}
    \end{table}


\subsection{Heating systems}
\label{subsec_C5_Heat_Cool}

\noindent An air-to-air heat pump (HP) driven multi-split system\footnote{The TRNSYS heat pump outdoor and indoor-fan units (\textit{Type} 221 and 222) are designed by Mr. Justin Tamasauskas as per the specifications are given in \Cref{C5_tab_Bungalow_Heating} (Email : \href{mailto:justin.tamasauskas@canada.ca}{justin.tamasauskas@canada.ca}).} is used as the primary heating unit of the house. The unit consists of an outdoor unit and three separate indoor fan units for the three zones. An electric baseboard (BB) heater is also available as a secondary heating unit for each zone. The HP heating performance depends on the outdoor temperature and may have lower coefficient of performance (COP) during the very cold days, hence for better reliability of the heating system the electric baseboards are used as the secondary units. The specifications are given in \Cref{C5_tab_Bungalow_Heating}. The required PLR for each indoor fan unit of the multi-split system, given by $PLR_{IN_{\text{demand}}}^i$, is optimized separately by the MPC. The total demanded heating capacity, $Q_{HP_{\!_{\text{demand}}}}$ is given by (\ref{C5_eqn_QHP_demand}). 


\vspace*{-0.1in}
\begin{align}
Q_{HP_{\!_{\text{demand}}}} & = \sum_{i = 1}^{z} \big(PLR_{IN_{\text{demand}}}^i \times Q_{IN_{\text{rated}}}^i \big) 
\label{C5_eqn_QHP_demand}
\end{align}

\noindent where,

\vspace*{0.1in}
\noindent\begin{tabularx}{\linewidth}{lX}
	$Q_{HP_{\!_{\text{demand}}}}$ & Total demanded HP heating capacity (kW).  \\
        $PLR_{IN_{\text{demand}}}^i$ & PLR requirement for the indoor fan unit for the $i^{th}$ zone. \\    
        $Q_{IN_{\text{rated}}}^i$ & Rated heating capacity of the indoor fan unit for the $i^{th}$ zone (kW).  \\
        $z$ & Total number of zones. Here, $z = 3$.
\end{tabularx}

\vspace*{0.2in}
The PLR of the outdoor unit, however, is adjusted based on the $Q_{HP_{\!_{\text{demand}}}}$ and the available rated heating capacity of the outdoor unit, ${Q_{HP_{\!_{\text{rated}}}}}$ (kW). The  ${Q_{HP_{\!_{\text{rated}}}}}$ and the electric power consumption,  $P_{HP_{\!_{\text{rated}}}}$ (kW) of the outdoor unit are functions of the indoor and outdoor temperature as given in \cite{Outdoor_unit}. PLR, effective heating capacity and electric power consumption of the outdoor unit are given by the set of equations in (\ref{C5_equ_HP_PLR_QHP_PHP}).

\begin{table}[pos=h!]
    \caption{Specifications for Heating and Cooling Units}
 \label{C5_tab_Bungalow_Heating}
{\setlength\extrarowheight{0.1pt}
    {\scriptsize\addtolength{\tabcolsep}{-6pt}
    \begin{tabular*}{\tblwidth}{@{} CCCCCCC@{} } 
    \toprule 
     \multicolumn{1}{ l }{ } &
     \multicolumn{1}{ p{0.1cm} }{} &
     \multicolumn{1}{ l }{\textbf{\  Parameters \ }} &
     \multicolumn{1}{ p{0.1cm} }{} &
     \multicolumn{1}{ l }{\textbf{\  Zones \ }}  &
     \multicolumn{1}{ p{0.1cm} }{} &
     \multicolumn{1}{ l }{\textbf{\  Rating \ }}  \\ \midrule 
     \multicolumn{1}{ l }{\ Outdoor Unit \ } &
     \multicolumn{1}{ p{0.1cm} }{} & 
     \multicolumn{1}{ l }{\ Total heating capacity \ } &
     \multicolumn{1}{ p{0.1cm} }{} &     
     \multicolumn{1}{ l }{\ - \ } &
     \multicolumn{1}{ p{0.1cm} }{} &     
     \multicolumn{1}{ l }{\ 45000 (BTU/h) \ } \\      
     \multicolumn{1}{ l }{\ \cite{Outdoor_unit} \ } &
     \multicolumn{1}{ p{0.1cm} }{} &
     \multicolumn{1}{ l }{\ Electric power (heating) \ } &
     \multicolumn{1}{ p{0.1cm} }{} &
     \multicolumn{1}{ l }{\ - \ } &
     \multicolumn{1}{ p{0.1cm} }{} &     
     \multicolumn{1}{ l }{\ 7200 W \ } \\ 	
     \multicolumn{1}{ l }{\  \ } &
     \multicolumn{1}{ p{0.1cm} }{} &      
     \multicolumn{1}{ l }{\ COP at -$8.33 \degree C$ ($17 \degree F$) \ } &
     \multicolumn{1}{ p{0.1cm} }{} &     
     \multicolumn{1}{ l }{\ - \ } &
     \multicolumn{1}{ p{0.1cm} }{} &     
     \multicolumn{1}{ l }{\ 2.85 \ } \\   \\ \\	
     \multicolumn{1}{ l }{\ Indoor Units  \ } &
     \multicolumn{1}{ p{0.1cm} }{} &      
     \multicolumn{1}{ l }{\ Total heating capacity \ } &
     \multicolumn{1}{ p{0.1cm} }{} &     
     \multicolumn{1}{ l }{\ B1, B2 \ } &
     \multicolumn{1}{ p{0.1cm} }{} &     
     \multicolumn{1}{ l }{\ 8700 (BTU/h) \ } \\      
     \multicolumn{1}{ l }{\ \cite{Indoor_unit_Bedrooms, Indoor_unit_LK} \ } &
     \multicolumn{1}{ p{0.1cm} }{} &
     \multicolumn{1}{ l }{\ Electric power (heating) \ } &
     \multicolumn{1}{ p{0.1cm} }{} &
     \multicolumn{1}{ l }{\ B1, B2 \ } &
     \multicolumn{1}{ p{0.1cm} }{} &     
     \multicolumn{1}{ l }{\ 545 W \ } \\     
     \multicolumn{1}{ l }{ } &
     \multicolumn{1}{ p{0.1cm} }{} &
     \multicolumn{1}{ l }{\ Heat air flow \ } &
     \multicolumn{1}{ p{0.1cm} }{} &
     \multicolumn{1}{ l }{\ B1, B2 \ } &
     \multicolumn{1}{ p{0.1cm} }{} &     
     \multicolumn{1}{ l }{\ 225 CFM \ } \\  
     \multicolumn{1}{ l }{\  \ } &
     \multicolumn{1}{ p{0.1cm} }{} &      
     \multicolumn{1}{ l }{\ COP at -$8.33 \degree C$ ($17 \degree F$) \ } &
     \multicolumn{1}{ p{0.1cm} }{} &     
     \multicolumn{1}{ l }{\ - \ } &
     \multicolumn{1}{ p{0.1cm} }{} &     
     \multicolumn{1}{ l }{\ 3.14 \ } \\
     \multicolumn{1}{ l }{ } &
     \multicolumn{1}{ p{0.1cm} }{} & 
     \multicolumn{1}{ l }{\ Total heating capacity \ } &
     \multicolumn{1}{ p{0.1cm} }{} &     
     \multicolumn{1}{ l }{\ LK \ } &
     \multicolumn{1}{ p{0.1cm} }{} &     
     \multicolumn{1}{ l }{\ 13600 (BTU/h) \ } \\      
     \multicolumn{1}{ l }{ } &
     \multicolumn{1}{ p{0.1cm} }{} &
     \multicolumn{1}{ l }{\ Electric power (heating) \ } &
     \multicolumn{1}{ p{0.1cm} }{} &
     \multicolumn{1}{ l }{\ LK \ } &
     \multicolumn{1}{ p{0.1cm} }{} &     
     \multicolumn{1}{ l }{\ 950 W \ } \\    
     \multicolumn{1}{ l }{ } &
     \multicolumn{1}{ p{0.1cm} }{} &
     \multicolumn{1}{ l }{\ Heat air flow \ } &
     \multicolumn{1}{ p{0.1cm} }{} &
     \multicolumn{1}{ l }{\ LK \ } &
     \multicolumn{1}{ p{0.1cm} }{} &     
     \multicolumn{1}{ l }{\ 454 CFM \ } \\  
     \multicolumn{1}{ l }{\  \ } &
     \multicolumn{1}{ p{0.1cm} }{} &      
     \multicolumn{1}{ l }{\ COP at -$8.33 \degree C$ ($17 \degree F$) \ } &
     \multicolumn{1}{ p{0.1cm} }{} &     
     \multicolumn{1}{ l }{\ - \ } &
     \multicolumn{1}{ p{0.1cm} }{} &     
     \multicolumn{1}{ l }{\ 2.10 \ } \\ \\ \\
     \multicolumn{1}{ l }{\ Baseboards \ } &
     \multicolumn{1}{ p{0.1cm} }{} & 
     \multicolumn{1}{ l }{\ Heating capacity \ } &
     \multicolumn{1}{ p{0.1cm} }{} &     
     \multicolumn{1}{ l }{\ B1 \ } &
     \multicolumn{1}{ p{0.1cm} }{} &     
     \multicolumn{1}{ l }{\ 2.25 kW \ } \\   
     \multicolumn{1}{ l }{\ \cite{BB} \ } &
     \multicolumn{1}{ p{0.1cm} }{} &
     \multicolumn{1}{ l }{ } &
     \multicolumn{1}{ p{0.1cm} }{} &
     \multicolumn{1}{ l }{\ B2 \ } &
     \multicolumn{1}{ p{0.1cm} }{} &     
     \multicolumn{1}{ l }{\ 2.00 kW \ } \\  
     \multicolumn{1}{ l }{ } &
     \multicolumn{1}{ p{0.1cm} }{} &
     \multicolumn{1}{ l }{ } &
     \multicolumn{1}{ p{0.1cm} }{} &
     \multicolumn{1}{ l }{\ LK \ } &
     \multicolumn{1}{ p{0.1cm} }{} &     
     \multicolumn{1}{ l }{\ 4.25 kW \ } \\ \bottomrule 
     \end{tabular*}}}
 \end{table}


\vspace*{-0.1in}
\begin{subequations}
\begin{flalign}
& PLR_{\!_{OU}} = 
	\begin{cases}
		\dfrac{Q_{HP_{\!_{\text{demand}}}}}{Q_{HP_{\!_{\text{rated}}}}}, & \ Q_{HP_{\!_{\text{demand}}}} <= Q_{HP_{\!_{\text{rated}}}}  \\
 		1, & \ \text{otherwise}
	\end{cases} 
\label{C5_equ_PLR_ou} & \\ 
& Q_{HP_{\!_{\text{OU}}}} = Q_{HP_{\!_{\text{rated}}}} \times PLR_{\!_{OU}}
\label{C5_equ_QHP_OU} \\ 
& P_{HP_{\!_{\text{OU}}}} = 
	\begin{cases}
		P_{HP_{\!_{\text{rated}}}} \times \big( 0.124 + 1.124 \times PLR_{\!_{OU}} \big), \\
		\hspace*{1.8in} PLR_{\!_{OU}} >= 0.4 \\
 		P_{HP_{\!_{\text{rated}}}} \times \big( 0.0109 + 0.9863 \times  PLR_{\!_{OU}}  \\
		\qquad - 2.3784 \times  PLR_{\!_{OU}}^2 + 4.8146 \times PLR_{\!_{OU}}^3 \big), \\ 
		\hspace*{2.1in} \text{otherwise}
	\end{cases} 
\label{C5_equ_PHP_OU} 
\end{flalign}
\label[equation]{C5_equ_HP_PLR_QHP_PHP} 
\end{subequations}	

\vspace*{-0.2in}
\noindent where,

\vspace*{0.1in}
\noindent\begin{tabularx}{\linewidth}{lX}
	$Q_{HP_{\!_{\text{rated}}}}$ & Rated heating capacity of the HP outdoor unit (kW). 
\end{tabularx}
\noindent\begin{tabularx}{\linewidth}{lX}     
        $P_{HP_{\!_{\text{rated}}}}$ & Rated power consumption by the HP outdoor unit (kW). \\
        $PLR_{\!_{OU}}$ & Effective PLR for the HP outdoor unit of the heat pump driven multi-split system. \\
        $Q_{HP_{\!_{\text{OU}}}}$ & Effective heating capacity of the HP outdoor unit (kW).  \\
        $P_{HP_{\!_{\text{OU}}}}$ & Effective power consumption by the HP outdoor unit (kW). 
\end{tabularx}

\vspace*{0.2in}
The relation between $PLR_{\!_{OU}}$ and $P_{HP_{\!_{\text{OU}}}}$ is adopted from the articles \cite{filliard2009performance, kegel2014integration}. The PLR of the indoor units are then adjusted by a heating capacity factor, $F_{\!_{\text{capacity}}}$ which is given by (\ref{C5_equ_F_cap}). The effective PLR for the indoor units are given by $PLR_{\!_{IN}}^i$ in (\ref{C5_PLR_in_adjusted}), for each zone.
\begin{subequations}
\begin{flalign}
F_{\!_{\text{capacity}}} & = 
	\begin{cases}
		1, & \ Q_{HP_{\!_{\text{demand}}}} <= Q_{HP_{\!_{\text{rated}}}}  \\
		\dfrac{Q_{HP_{\!_{\text{rated}}}}}{Q_{HP_{\!_{\text{demand}}}}}, & \ \text{otherwise}
	\end{cases} 
		\label{C5_equ_F_cap} & \\ 
PLR_{\!_{IN}}^i & = PLR_{IN_{\text{demand}}}^i \times F_{\!_{\text{capacity}}}
\label{C5_PLR_in_adjusted} 
\end{flalign}
\label[equation]{C5_equ_HP_PLR_in} 
\end{subequations}				
	
\vspace*{-0.1in}   
\noindent where,

\vspace*{0.1in}
\noindent\begin{tabularx}{\linewidth}{lX}
        $F_{\!_{\text{capacity}}}$ & Heating capacity factor for the indoor units of the multi-split system.  \\
        $PLR_{\!_{IN}}^i$ & Effective PLR for the indoor fan unit in the $i^{th}$ zone.  
\end{tabularx}

\vspace*{0.1in}	
Equations (\ref{C5_eqn_QHP_demand}), (\ref{C5_equ_HP_PLR_QHP_PHP}) and (\ref{C5_equ_HP_PLR_in}) ensure that the total heating capacity of the individual fan units, in the zones, never exceeds the rated capacity of the outdoor unit of the multi-split system. While the MPC only optimizes the $PLR_{IN_{\text{demand}}}^i$ as the control variable, the adjustments of these PLR values for the indoor and outdoor units, are calculated as a part of the optimization process. $PLR_{\!_{OU}}$ is a function of  $PLR_{IN_{\text{demand}}}^i$ as given by \cref{C5_eqn_QHP_demand,C5_equ_PLR_ou}. Hence, by the end of the iterative calculations, constrained optimal values of $PLR_{\!_{IN}}^i$ and $PLR_{\!_{OU}}$ are available. In the rest of the text, $PLR_{\!_{IN}}$ is referred to as the HP control variable optimized by the MPC.

The baseboards, on the other hand, are assumed to convert 100\% of the consumed electricity to heating energy. Unlike the radiant floor units, discussed in the previous chapters, the heat pump driven multi-split systems have a lower installation cost and a faster response time. The only drawback to consider is that the performance of the heat pump depends on the outdoor and indoor temperatures, among other factors. Specially, since no thermal storage is considered here, when the heat pump is operating at a lower COP, the MPC utilizes the baseboards to provide the heating energy required to maintain the indoor operative temperature within the desired range. Thus, the baseboards perform as a stand-by heating option for the control scenarios discussed in the following. 
	
\subsection{PV panels (PV), home battery (HB) and grid (G): sizing and specifications}
\label{C5_subsec_PV_HB_Grid}

\noindent Renewable solar energy generation and its storage using a home battery are considered here as a part of the BEMS. The goal is to maximize the generation and self-utilization of the green energy and optimally use the battery storage unit to reduce the energy cost. The MPC, on the one hand, optimizes the heating energy usage and simultaneously manages the PV generation and utilization of the battery storage to distribute the energy consumption in such a way that the consumer can benefit from the variable ToU electricity rates during the day. 

The two-way grid is considered as an infinite source or sink as necessary. The battery can act as a power source (discharging) or a load (charging) at a given time step while the PV is always a source and the house is a load. The MPC determines the optimal amount of power flow, at each time step, from the sources to the loads. At every time step, the grid supplies the required power when the PV and the battery are unable to meet the demand. Any excess generated power, at a given instant, is sent back to the grid. The MPC energy management strategy is discussed in detail in \Cref{C5_sec_Prob}. The specifications of the PV and the home battery are given in the following.
	

\subsubsection{Solar panels}
\label{C5_subsubsec_PV}

\noindent The specifications for the PV panels are given in \Cref{C5_tab_Bungalow_PV}. A simple glazed PV panel, (TRNSYS \textit{Type 562d} module) is used to estimate the solar power generation using the solar radiation data given by the TMY2 weather file. The default specifications for the said PV module remained unchanged, as shown in the table, except the surface area. It is estimated to be $32.7 \ m^2$, assuming the average size of the PV system to be 6 kW \cite{PV_panel}. The residential rooftop panels generally have 60 cells, with each small cell having $6'' \times 6''$ dimension and a 60-cell panel generates about 300 W. Thus, a set of 20 average-sized panels, generating 6 kW solar power, has an approximate size of $27' \times 13'$ which is equivalent to 32.7 m$^2$ \cite{PV_panel}. The South facing slanted roof, where the panels are installed, has an area of 53.74 m$^2$. Hence, sufficient spacing can be provided between the panels.

\begin{table}[pos=h!]
    \caption{Specifications for the solar panels}
     \label{C5_tab_Bungalow_PV}
{\setlength\extrarowheight{0.1pt}
    {\scriptsize\addtolength{\tabcolsep}{-6pt}
    \begin{tabular*}{\tblwidth}{@{} CCCCC@{} }
    \toprule 
     \multicolumn{1}{ l }{ } &
     \multicolumn{1}{ p{1cm} }{} &
     \multicolumn{1}{ l }{\textbf{\  Parameters \ }} &
     \multicolumn{1}{ p{0.5cm} }{} &
     \multicolumn{1}{ l }{\textbf{\  Rating \ }}  \\ \midrule 
     \multicolumn{1}{ l }{\ Solar panels \ } &
     \multicolumn{1}{ p{0.5cm} }{} & 
     \multicolumn{1}{ l }{\ Total surface area \ } &
     \multicolumn{1}{ p{0.5cm} }{} &     
     \multicolumn{1}{ l }{\ 32.7 m$^2$ \ } \\   
     \multicolumn{1}{ l }{\ \cite{PV_TRNSYS_ref, PV_panel}} &
     \multicolumn{1}{ p{0.5cm} }{} &
     \multicolumn{1}{ l }{\ Reference PV efficiency \ } &
     \multicolumn{1}{ p{0.5cm} }{} &     
     \multicolumn{1}{ l }{\ 14.88 \% \ } \\  
     \multicolumn{1}{ l }{\  \ } &
     \multicolumn{1}{ p{0.5cm} }{} &
     \multicolumn{1}{ l }{\ Reference temperature \ } &
     \multicolumn{1}{ p{0.5cm} }{} &     
     \multicolumn{1}{ l }{\ $25 \degree C$ \ }  \\  
     \multicolumn{1}{ l }{\  \ } &
     \multicolumn{1}{ p{0.5cm} }{} &
     \multicolumn{1}{ l }{\ Absorptance \ } &
     \multicolumn{1}{ p{0.5cm} }{} &     
     \multicolumn{1}{ l }{\ 0.855 \ } \\ \bottomrule
     \end{tabular*}}}
\end{table}

The operations of the PV panel is affected by different factors of the ambient conditions, e.g., solar irradiance level, ambient temperature, wind speed, dirt/dust etc. \cite{meral2011review}. The efficiency of the PV panel is a function of the reference PV efficiency, the reference and actual temperatures the solar cell, total reference and actual incident solar radiation \cite{PV_TRNSYS_ref}. Since, an estimation of the generated power, based on a given weather condition, suffices the requirements of this particular study, a simplified model given by (\ref{equ_C5_PPV}) is used for the simulation. 
\begin{equation}
P_{\!_{PV}} = \eta_{\!_{PV}} \times SI \times A_{\!_{PV}} \times (\tau\alpha)_n \times \eta_{\!_{RI}}
\label{equ_C5_PPV}
\end{equation}

\noindent where,

\vspace*{0.1in}
\noindent\begin{tabularx}{\linewidth}{lX}
	$P_{\!_{PV}}$ & Total solar power generation (kW). \\  
        $SI$ & Total solar irradiance (W/m$^2$). \\      
        $A_{\!_{PV}}$ & Total surface area of the PV panels (m$^2$).  \\
        $\eta_{\!_{PV}}$ & Efficiency of the PV panels. \\ 
        $\eta_{\!_{RI}}$ & Efficiency of the Regulator - Inverter unit. \\
        $(\tau\alpha)_n$ & Transmittance - absorptance product of the PV cover for solar radiation at a normal angle.
\end{tabularx} 

\vspace*{0.1in}
The efficiency of the regulator-inverter unit is set to 78\%, equal to the default value of \textit{Type 48} module in TRNSYS \cite{RI_TRNSYS_ref}. The data for solar irradiance ($SI$) and the PV efficiency ($\eta_{\!_{PV}}$) are recorded from TRNSYS for the TMY2 - Toronto weather data. The normal solar transmittance ($\tau_n$) of the glass cover value is set to 0.86. Assuming the absorption factor, in \Cref{C5_tab_Bungalow_PV}, as the normal absorption factor ($\alpha_n$), the $(\tau\alpha)_n$ is approximated as $(\tau\alpha)_n = 1.01 \times \tau_n \times \alpha_n = 0.74$ \cite{duffie2013solar}. With this rough approximations of the PV panel parameter specifications, the power generation simulated by the model in (\ref{equ_C5_PPV}) generates comparable results with respect to that of the \textit{Type 562d} module in TRNSYS. 

%

It is worth mentioning that the simulation of the PV panels, is performed entirely within the MATLAB environment and the \textit{Type 562d} for PV panels and \textit{Type 48} for the regulator-inverter module in TRNSYS are only used to derive the simplified model and its validation.


\subsubsection{Home-battery}
\label{C5_subsubsec_HB}

\noindent The specifications for the home battery is given in \Cref{C5_tab_Bungalow_Bat}. The specifications are similar to the Tesla Powerwall \cite{HB_powerwall}. Except for the base case simulation scenario where for a rule based control strategy the minimum allowable state of charge (SOC) is restricted to 50\%, in general, the SOC is allowed to vary between 10\% - 90\%.

\begin{table}[pos=h!]
    \caption{Specifications for the home battery}
     \label{C5_tab_Bungalow_Bat}
{\setlength\extrarowheight{0.1pt}
    {\scriptsize\addtolength{\tabcolsep}{-6pt}
    \begin{tabular*}{\tblwidth}{@{} CCC@{} }
    \toprule 
     \multicolumn{1}{ l }{ } &
     \multicolumn{1}{ l }{\textbf{\  Parameters \ }} &
     \multicolumn{1}{ l }{\textbf{\  Rating \ }}  \\ \midrule
     \multicolumn{1}{ l }{\ Home-battery \ } &
     \multicolumn{1}{ l }{\ Nominal AC voltage \ } &
     \multicolumn{1}{ l }{\ 120 / 240 V \ } \\   
     \multicolumn{1}{ l }{\ \cite{HB_powerwall}} &
     \multicolumn{1}{ l }{\ Grid frequency \ } &
     \multicolumn{1}{ l }{\ 60 Hz \ } \\  
     \multicolumn{1}{ l }{\  \ } &
     \multicolumn{1}{ l }{\ Rated energy \ } &
     \multicolumn{1}{ l }{\ 14 kWh \ }  \\  
     \multicolumn{1}{ l }{\  \ } &
     \multicolumn{1}{ l }{\ Usable energy, $E_{B_{max}}$ \ } &  
     \multicolumn{1}{ l }{\ 13.5 kWh \ } \\ 
     \multicolumn{1}{ l }{\  \ } &
     \multicolumn{1}{ l }{\ Rated power \ } &
     \multicolumn{1}{ l }{\ 5 kW \ } \\ \bottomrule 
     \end{tabular*}}}
\end{table}

\noindent The SOC of the home battery is calculated in a discrete time frame at every simulation time step \cite{mazzeo2018energy}. The update equations are given by \cref{equ_C5_E_HB,equ_C5_SOC_HB}.
\begin{flalign}
E_{\!_B}(k + 1) & = 
	\begin{cases}
		E_{\!_B}(k) + P_{\!_B} \times \eta_{\!_B} \times \Delta k, \ P_{\!_B} > 0, \ \text{charging} \\
		E_{\!_B}(k) + P_{\!_B} \times \eta_{\!_B} \times \Delta k, \ P_{\!_B} < 0, \ \text{discharging}
	\end{cases}
\label{equ_C5_E_HB} & \\ 
SOC_{\!_B}(k) & = \dfrac{E_{\!_B}(k) \times 100}{E_{B_{max}}} 
\label{equ_C5_SOC_HB} \\
SOC_{\!_{B_{min}}} \ & \ \leq \ SOC_{\!_B}(k) \ \leq \ SOC_{\!_{B_{max}}} 
\label{equ_C5_SOC_range}
\end{flalign}

\noindent where,

\vspace*{0.1in}
\noindent\begin{tabularx}{\linewidth}{lX}
        $E_{\!_B}$ & Stored energy in the home battery at a given time step (kWh). \\  
        $P_{\!_B}$ & Amount of power sent to or drawn from the home battery (kW). \\        
        $\eta_{\!_{B}}$ & Home-battery efficiency = 95\%. \\ 
        $\Delta k$ & Time interval = $T_s / 3600$ (h). \\
        $T_s$ & Simulation time step (s).  \\
        $SOC_{\!_B}$ & SOC of the home battery at a given time step (\%). \\
        $SOC_{\!_{B_{min}}}$ & Minimum allowable SOC for the home battery. \\
        $SOC_{\!_{B_{max}}}$ & Maximum allowable SOC for the home battery.
\end{tabularx}


\section{Control-oriented model identification and validation}
\label{C5_Sys_ID_Val}

\begin{figure}[pos=h!]
    \centering
    \includegraphics[scale = 0.8]{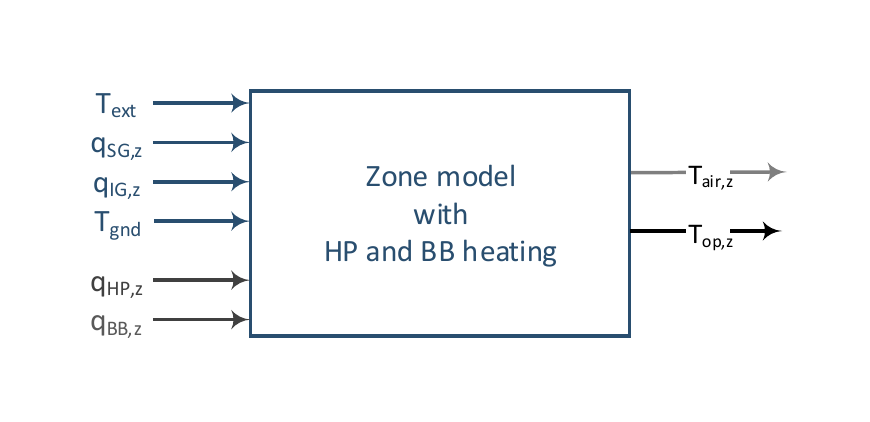}
\vspace*{-0.2in}
    \caption{Identified system model}
\vspace*{-0.2in}
    \label{C5_fig_Bungalow_model}
\end{figure}

\noindent The control-oriented MATLAB models for the three zones, namely, the primary bedroom, secondary bedroom and the living-kitchen area (\Cref{fig_C5_Bungalow}) are identified separately as linear time-invariant (LTI) state space sub-blocks as shown in \Cref{C5_fig_Bungalow_model}. \\

\noindent \textbf{System inputs: } 
\begin{itemize}[noitemsep]
	\item External uncontrolled or disturbance inputs : 
		\begin{itemize}[noitemsep]
			\item $T_{ext}$ : Outdoor temperature ($\degree C$). 
			\item $q_{\!_{SG,z}}$ : Solar gain (kW) for $z^{th}$ zone.
			\item $q_{\!_{IG,z}}$ : Internal thermal gains (kW) for $z^{th}$ zone. It represents the overall heat gain from the occupants and appliances present in the house. 
			\item $T_{gnd}$ : Ground temperature ($\degree C$). 
		\end{itemize}
\item Heating inputs:
	\begin{itemize}[noitemsep]
		\item $q_{\!_{HP,z}}$ : Power inputs for the multi-split heat pump unit (kW) for $z^{th}$ zone.
		\item $q_{\!_{BB,z}}$ : Power inputs for the baseboards (kW) for $z^{th}$ zone.
	\end{itemize}
\end{itemize}

\vspace*{-0.2in}

\noindent Open loop system identification is performed to generate the simplified LTI house model. The MPC uses this control-oriented model to predict the system outputs during the heating season and accordingly optimizes the control inputs.  Periodic pulse waves of 50\% duty cycle, having amplitudes equal to the rated HP and BB heating capacity  (\Cref{C5_tab_Bungalow_Heating}) are supplied to respective zones. \\

\noindent \textbf{System outputs: } 
\vspace*{-0.1in}
\begin{itemize}[noitemsep]
	\item $T{\!_{op,z}}$ : Operative temperature ($\degree C$) for $z^{th}$ zone.
	\item $T_{\!_{air,z}}$ : Air temperature ($\degree C$) for $z^{th}$ zone.
\end{itemize}

\noindent For system identification, the TMY2 Toronto weather data from January to the middle of March is used. The model validation is performed with the weather dataset for the same location, from the middle of March to the end of April. Distribution of the validation errors for the different zones vary wiithin $\pm 2 \degree C$ for close to 70\% of the time.

\section{Uncertainties in the disturbance inputs: weather inputs and internal gains}
\label{C5_sec_uncertainties}

\noindent Forecast data is available for $T_{ext}$ and $q_{\!_{SG}}$. But, as mentioned in \cite{candanedo2018preliminary}, using the available forecast data from weather stations in the simulation studies is not straightforward and there is no standard method of approximating the effects of forecast related uncertainties. For the simulation experiments presented in this article, experiment based recommendations suggested in \cite{candanedo2018preliminary} are adopted. A detailed description of the different noise vectors generated as uncertainties for the uncontrolled system inputs is described in this section.

Here, the standard TRNSYS TMY2 dataset for Toronto is used as the actual or real-time data and will be referred to like the same in the following. The noise vectors are generated for every hour \cite{candanedo2018preliminary} and linear interpolation is used between successive data points to generate a noise vector that matches the simulation time step of 3.75 min. These noise vectors are then added to the TMY2 dataset to generate the forecast data.
 


\subsection{Forecast and actual dataset for outdoor temperature}
\label{C5_subsec_uncertainty_Text}

\noindent As suggested in \cite{candanedo2018preliminary}, two vectors of normally distributed random variables are considered to be representing, respectively, the 6 hour-ahead and the 1 hour-ahead forecast noise for the outdoor temperature, $T_{ext}$. The mean and standard deviation of the prediction error considered for the 6 hour-ahead forecast are $0.04 \degree C$ and $2.2 \degree C$ respectively. For the 1 hour-ahead forecast, the prediction accuracy is assumed to be improved with a narrower standard deviation of $1.2 \degree C$. 



\subsection{Forecast and actual dataset for solar radiation on the PV and shortwave solar heat gain through windows}
\label{C5_subsec_uncertainty_solar}


\noindent 1 hour, 6 hour and 48 hour-ahead forecasts for solar radiation are studied in \cite{candanedo2018preliminary} based on available weather dataset from Environment Canada. In the article, an estimated forecast error is defined by Gaussian noise with mean of 17 W/m$^2$ and  standard deviation of 167 W/m$^2$ for the 6 hour-ahead predictions. The 1 hour-ahead prediction, is a Gaussian noise with the same mean but a smaller standard deviation of 100 W/m$^2$. As indicated in the article mentioned above, these are the prediction estimations for the solar radiation due $45 \degree$ South and hence it is suitable for the solar radiation on the installed PV panels.

To generate the forecast error for the indoor solar heat gain due to the shortwave radiation through the windows, Gaussian noises with similar mean and standard deviation are used as the base for the 6 hour and 1 hour-ahead predictions. These base noise vectors are then weighted by the following three factors:

\vspace*{0.1in}
\noindent\begin{tabularx}{\linewidth}{lX}
$Q_{sol_{factor}}$ & The relative solar heat gain factors \cite{Qsol_heat_gain_factor}, for the vertical fa\c{c}ades at the different orientation of the building and at a different time of the day, with respect to South fa\c{c}ade. \\
$A_w$ & Total window area (\Cref{C5_tab_Bungalow_Specs}) on each fa\c{c}ade corresponding to a zone (m$^2$). \\
G-value & The solar energy transmittance of the windows. The G-value of the window glass is assumed to be 0.376. 
\end{tabularx}

\vspace*{0.1in}
\noindent The Winter (corresponding to January $21^{st}$) solar heat gain factors at $43 \degree N$ latitude for Toronto, are considered as given in \cite{Qsol_heat_gain_factor}. These factors are normalized with respect to the heat gain on the South fa\c{c}ade and are listed in \Cref{C5_tab_Bungalow_solar_gain}.

\begin{table}[pos=h!]
    \caption{Relative solar heat gain factors with respect to the heat gain on South fa\c{c}ade}
     \label{C5_tab_Bungalow_solar_gain}
{\setlength\extrarowheight{0.1pt}
    {\scriptsize\addtolength{\tabcolsep}{-6pt}
    \begin{tabular*}{\tblwidth}{@{} CCCCCCCCCCCC@{} } 
    \toprule 
     \multicolumn{1}{ c }{\textbf{\ Time \ }} &
     \multicolumn{1}{ p{0.5cm} }{} &
     \multicolumn{1}{ c }{\textbf{\ North \ }} &
     \multicolumn{1}{ p{0.5cm} }{} &
     \multicolumn{1}{ c }{\textbf{\ East \ }} &
     \multicolumn{1}{ p{0.5cm} }{} &
     \multicolumn{1}{ c }{\textbf{\ South\ }} &
     \multicolumn{1}{ p{0.5cm} }{} &
     \multicolumn{1}{ c }{\textbf{\ West \ }} &
     \multicolumn{1}{ p{0.5cm} }{} &
     \multicolumn{1}{ c }{\textbf{\  Time \ }} \\
     \multicolumn{1}{ c }{\textbf{\ AM \ }} &
     \multicolumn{1}{ p{0.5cm} }{} &
     \multicolumn{1}{ c }{\textbf{\   \ }} &
     \multicolumn{1}{ p{0.5cm} }{} &
     \multicolumn{1}{ c }{\textbf{\   \ }} &
     \multicolumn{1}{ p{0.5cm} }{} &
     \multicolumn{1}{ c }{\textbf{\  \ }} &
     \multicolumn{1}{ p{0.5cm} }{} &
     \multicolumn{1}{ c}{\textbf{\   \ }}  &
     \multicolumn{1}{ p{0.5cm} }{} &
     \multicolumn{1}{ c }{\textbf{\ PM \ }} \\ \midrule 
     \multicolumn{1}{ c }{\ 8 \ } &
     \multicolumn{1}{ p{0.5cm} }{} &
     \multicolumn{1}{ c }{\ 0.053 \ } &
     \multicolumn{1}{ p{0.5cm} }{} &
     \multicolumn{1}{ c }{\ 1.474 \ } &
     \multicolumn{1}{ p{0.5cm} }{} &
     \multicolumn{1}{ c }{\ 1 \ } &
     \multicolumn{1}{ p{0.5cm} }{} &
     \multicolumn{1}{ c }{\ 0.053 \ } &
     \multicolumn{1}{ p{0.5cm} }{} &
     \multicolumn{1}{ c }{\ 4 \ } \\ 
     \multicolumn{1}{ c }{\ 9 \ } &
     \multicolumn{1}{ p{0.5cm} }{} &
     \multicolumn{1}{ c }{\ 0.066 \ } &
     \multicolumn{1}{ p{0.5cm} }{} &
     \multicolumn{1}{ c }{\ 0.940 \ } &
     \multicolumn{1}{ p{0.5cm} }{} &
     \multicolumn{1}{ c }{\ 1 \ } &
     \multicolumn{1}{ p{0.5cm} }{} &
     \multicolumn{1}{ c }{\ 0.066 \ } &
     \multicolumn{1}{ p{0.5cm} }{} &
     \multicolumn{1}{ c }{\ 3 \ } \\ 
     \multicolumn{1}{ c }{\ 10 \ } &
     \multicolumn{1}{ p{0.5cm} }{} &
     \multicolumn{1}{ c }{\ 0.067 \ } &
     \multicolumn{1}{ p{0.5cm} }{} &
     \multicolumn{1}{ c }{\ 0.565 \ } &
     \multicolumn{1}{ p{0.5cm} }{} &
     \multicolumn{1}{ c }{\ 1 \ } &
     \multicolumn{1}{ p{0.5cm} }{} &
     \multicolumn{1}{ c }{\ 0.067 \ } &
     \multicolumn{1}{ p{0.5cm} }{} &
     \multicolumn{1}{ c }{\ 2 \ } \\ 
     \multicolumn{1}{ c }{\ 11 \ } &
     \multicolumn{1}{ p{0.5cm} }{} &
     \multicolumn{1}{ c }{\ 0.070 \ } &
     \multicolumn{1}{ p{0.5cm} }{} &
     \multicolumn{1}{ c }{\ 0.240 \ } &
     \multicolumn{1}{ p{0.5cm} }{} &
     \multicolumn{1}{ c }{\ 1 \ } &
     \multicolumn{1}{ p{0.5cm} }{} &
     \multicolumn{1}{ c }{\ 0.070 \ } &
     \multicolumn{1}{ p{0.5cm} }{} &
     \multicolumn{1}{ c }{\ 1 \ } \\ 
     \multicolumn{1}{ c }{\ 12 \ } &
     \multicolumn{1}{ p{0.5cm} }{} &
     \multicolumn{1}{ c }{\ 0.067 \ } &
     \multicolumn{1}{ p{0.5cm} }{} &
     \multicolumn{1}{ c }{\ 0.075 \ } &
     \multicolumn{1}{ p{0.5cm} }{} &
     \multicolumn{1}{ c }{\ 1 \ } &
     \multicolumn{1}{ p{0.5cm} }{} &
     \multicolumn{1}{ c }{\ 0.075 \ } &
     \multicolumn{1}{ p{0.5cm} }{} &
     \multicolumn{1}{ c }{\ 12 \ } \\ \bottomrule 
     \end{tabular*}}}
\end{table}

\noindent The Gaussian noise generated for the indoor solar heat gain on a given fa\c{c}ade is given by the equation:
\begin{equation}
   \overline{Q}_{sol_{noise}} = \Delta \overline{q}_{\!_{SG}}  \times \text{G-value} \times A_w \times Q_{sol_{factor}}
    \label{C5_eqn_indoor_Qsg}
\end{equation}

\noindent where,

\vspace*{0.1in}
\noindent\begin{tabularx}{\linewidth}{lX}
       $\Delta \overline{q}_{\!_{SG}}$ & Base Gaussian noise vector for solar heat gain per unit area due South (W/m$^2$). \\
       \end{tabularx}  
\begin{tabularx}{\linewidth}{lX}
      $\overline{Q}_{sol_{noise}}$ & Net solar heat gain noise vector on a given fa\c{c}ade (W). 
\end{tabularx}        

\vspace*{0.1in}
\noindent The solar heat gain noise, $\overline{Q}_{sol_{noise}}$ is generated separately for each fa\c{c}ade of the house using (\ref{C5_eqn_indoor_Qsg}), and then the total noise for each zone is added to the real-time data to generate the forecast to be used by the MPC. 


\subsection{Forecast and actual dataset for ground temperature and internal heat gain}
\label{C5_subsec_uncertainty_Tgnd_Qig}


\noindent The variation in the ground surface temperature is $\pm 0.5 \degree C$ \cite{beltrami2015ground}. Gaussian noise of $0 \degree C$ mean and $0.17 \degree C$ standard deviation is generated which gives the required range of prediction error for the ground temperature, $T_{gnd}$. 

The internal heat gain is considered to be generated by the occupants, lights and electrical equipment. The schedules for occupancy and electricity usage for lights and appliances are pre-scheduled in the TRNSYS house. Hence, a Gaussian percentage deviation with zero mean and an approximate maximum deviation of $\pm 30\%$ is considered. 

%
  
 
\section{Problem formulation and methodology}
\label{C5_sec_Prob}

\noindent The goal here is to combine the control mechanism of two major building operations. The comfort goal focuses on meeting a predefined temperature range while reducing in the energy cost. Optimal usage of the heating units, thus, becomes an integral part of the comfort optimization. Judicious utilization of the energy efficient heat pump and benefiting from the off-peak ToU rates during the day by utilizing the energy storage unit, prove to be the two key factors implemented by the MPC for the comfort optimization \cite{Seal_2019,Seal_thesis}. 

The self-generated electricity from the roof-top PV panels can be used instantaneously to supply loads or can be stored in the home battery for future usage in a cost-effective way. The heating system is considered as a part of the house electricity load and thus the heating control system and the energy management system are intertwined to study a more centralized control scenario. The MPC strategy proposed here acts as a central control system to handle the heating systems as well as the PV generation, the battery usage and also the interaction with the grid. The base case scenario, defined in this case study, includes an MPC controlled heating system and a rule based energy management system.


\subsection{Comfort bounds and setpoint temperature}
\label{C5_subsec_setpoint_heating}
	
\begin{table*}[pos=h!]
    \caption{Summary of control variables for cost and comfort optimization for a house with PV panels and home battery}
     \label{C5_tab_PV_opt_var}
{\setlength\extrarowheight{0.1pt}
    {\scriptsize\addtolength{\tabcolsep}{-6pt}
    \centering
    \begin{tabular}{cccccccccccl}
    \toprule 
     \multicolumn{1}{ c }{\textbf{\ Variable \ }} &
     \multicolumn{1}{ p{0.5cm} }{} &
     \multicolumn{1}{ c }{\textbf{\ Range \ }} &
     \multicolumn{1}{ p{0.5cm} }{} &
     \multicolumn{1}{ c }{\textbf{\ Initial guess \ }} &
     \multicolumn{1}{ p{0.5cm} }{} &
     \multicolumn{1}{ c }{\textbf{\ Unit \ }} &
     \multicolumn{1}{ p{0.5cm} }{} &
     \multicolumn{1}{ c }{\textbf{\ Count \ }} &
     \multicolumn{1}{ p{0.5cm} }{} &
     \multicolumn{1}{ l }{\textbf{\ Description \ }} \\ \midrule 
     \multicolumn{1}{ c }{\ $PLR_{\!_{IN}}$ \ } &
     \multicolumn{1}{ p{0.5cm} }{} &
     \multicolumn{1}{ c }{\ [0, 1] \ } &
     \multicolumn{1}{ p{0.5cm} }{} &
     \multicolumn{1}{ c }{\ 0.2 \ } &
     \multicolumn{1}{ p{0.5cm} }{} &
     \multicolumn{1}{ c }{\ - \ } &
     \multicolumn{1}{ p{0.5cm} }{} &
     \multicolumn{1}{ c }{\ 3 \ } &
     \multicolumn{1}{ p{0.5cm} }{} &
     \multicolumn{1}{ l }{\ Part load ratio for each HP indoor fan unit. \ } \\ 
     \multicolumn{1}{ c }{\ $F_{\!_{BB}}$ \ } &
     \multicolumn{1}{ p{0.5cm} }{} &
     \multicolumn{1}{ c }{\ [0, 1] \ } &
     \multicolumn{1}{ p{0.5cm} }{} &
     \multicolumn{1}{ c }{\ 10 \ } &
     \multicolumn{1}{ p{0.5cm} }{} &
     \multicolumn{1}{ c }{\ \% \ } &
     \multicolumn{1}{ p{0.5cm} }{} &
     \multicolumn{1}{ c }{\ 3 \ } &
     \multicolumn{1}{ p{0.5cm} }{} &
     \multicolumn{1}{ l }{\ Fraction of the rated baseboard heat input, $q_{\!_{BB_{rated}}}$, for each zone, \ } \\ 
     \multicolumn{1}{ c }{\  \ } &
     \multicolumn{1}{ p{0.5cm} }{} &
     \multicolumn{1}{ c }{\  \ } &
     \multicolumn{1}{ p{0.5cm} }{} &
     \multicolumn{1}{ c }{\ \ } &
     \multicolumn{1}{ p{0.5cm} }{} &
     \multicolumn{1}{ c }{\  \ } &
     \multicolumn{1}{ p{0.5cm} }{} &
     \multicolumn{1}{ c }{\  \ } &
     \multicolumn{1}{ p{0.5cm} }{} &
     \multicolumn{1}{ l }{\  as listed in \Cref{C5_tab_Bungalow_Heating}. \ } \\ 
     \multicolumn{1}{ c }{\ $P_{\!_{B}}$ \ } &
     \multicolumn{1}{ p{0.5cm} }{} &
     \multicolumn{1}{ c }{\ [-5, 5] \ } &
     \multicolumn{1}{ p{0.5cm} }{} &
     \multicolumn{1}{ c }{\ 0 \ } &
     \multicolumn{1}{ p{0.5cm} }{} &
     \multicolumn{1}{ c }{\ kW \ } &
     \multicolumn{1}{ p{0.5cm} }{} &
     \multicolumn{1}{ c }{\ 1 \ } &
     \multicolumn{1}{ p{0.5cm} }{} &
     \multicolumn{1}{ l }{\ Charging and discharging power for the home battery. \ } \\ 
     \multicolumn{1}{ c }{\ $F_{\!_{PV \rightarrow H}}$ \ } &
     \multicolumn{1}{ p{0.5cm} }{} &
     \multicolumn{1}{ c }{\ [0, 1] \ } &
     \multicolumn{1}{ p{0.5cm} }{} &
     \multicolumn{1}{ c }{\ 0.3 \ } &
     \multicolumn{1}{ p{0.5cm} }{} &
     \multicolumn{1}{ c }{\ \% \ } &
     \multicolumn{1}{ p{0.5cm} }{} &
     \multicolumn{1}{ c }{\ 1 \ } &
     \multicolumn{1}{ p{0.5cm} }{} &
     \multicolumn{1}{ l }{\ Fraction of $P_{\!_{PV}}$ supplied from the PV to the house.  \ } \\ 
     \multicolumn{1}{ c }{\ $F_{\!_{B \rightarrow H}}$ \ } &
     \multicolumn{1}{ p{0.5cm} }{} &
     \multicolumn{1}{ c }{\ [0, 1] \ } &
     \multicolumn{1}{ p{0.5cm} }{} &
     \multicolumn{1}{ c }{\ 0.6 \ } &
     \multicolumn{1}{ p{0.5cm} }{} &
     \multicolumn{1}{ c }{\ \% \ } &
     \multicolumn{1}{ p{0.5cm} }{} &
     \multicolumn{1}{ c }{\ 1 \ } &
     \multicolumn{1}{ p{0.5cm} }{} &
     \multicolumn{1}{ l }{\ Fraction of $P_{\!_{B}}$ supplied from the home battery to the house. \ } \\ \bottomrule 
     \end{tabular}}}
 \end{table*}
 
\begin{figure}[pos=h!]
    \centering
    \includegraphics[scale = 0.6]{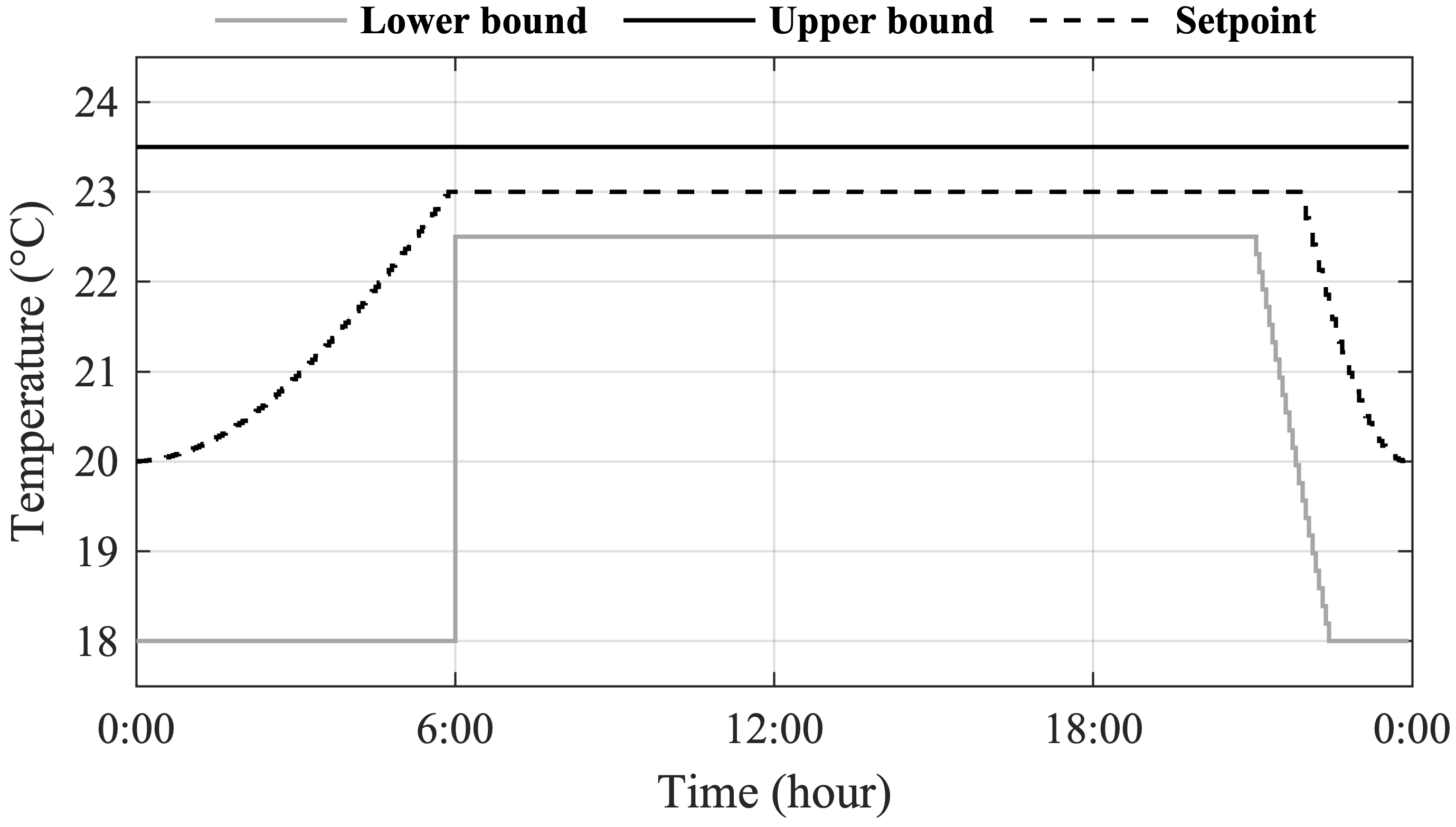}
    \caption{Setpoint temperature and bounds on the indoor operative temperature}
    \label{fig_C5_setpoint} 
  \end{figure}

 \noindent The temperature bounds considered here are shown in \Cref{fig_C5_setpoint}. During the day, the zone operative temperatures are maintained around $23 \degree C$ with an acceptable allowance of $\pm 0.5 \degree C$.  For the night, more flexible bounds are set as indicated in the figure.  In addition to the upper ($T_{ub}$) and lower ($T_{lb}$) acceptable deviation limits, a setpoint ($T_{set}$) is also considered for the simulation experiments (\Cref{fig_C5_setpoint}). 
 
It has been identified in \cite{Seal_2019,Seal_thesis}, that overheating is often caused by the MPC to achieve a suitable trade-off between cost and comfort, specially in the absence of forced cooling. To reduce this effect of overheating, the distance of the operative temperature, $T_{op}$ from $T_{set}$  is penalized in the objective function in addition to penalizing the deviation from the bounds. The temperature profile for $T_{set}$ is chosen to maintain approximately the average of the upper and lower bounds, while it also follows the natural heating dynamics of the system.

The penalty on comfort violation used in the MPC optimization is defined in  \cref{equ_C5_cntrl_tracking_error_Ebound,equ_C5_cntrl_tracking_error_Eset}. Two different non-negative tracking errors are taken into consideration. The $\overline{e}_{bound}^i : \mathbb{R}^{L} \times \mathbb{R}^{L} \times \mathbb{R}^{L} \rightarrow \mathbb{R}^{L}$ measures the deviation of $T_{op}$ from the upper and lower temperature bounds (\ref{equ_C5_cntrl_tracking_error_Ebound}). The  $\overline{e}_{set}^i : \mathbb{R}^{L} \times \mathbb{R}^{L} \rightarrow \mathbb{R}^{L}$ in (\ref{equ_C5_cntrl_tracking_error_Eset}), on the other hand, is the absolute measure of the distance of $T_{op}$ from $T_{set}$.

 \vspace*{-0.12in}
 \begin{flalign}
    \overline{e}_{set}^i (\overline{T}_{op}^{i}, \overline{T}_{set}^{i}) & =  \big| \ \overline{T}^i_{op} - \overline{T}^i_{set} \ \big| 
    \label{equ_C5_cntrl_tracking_error_Eset} & \\ 
    \overline{e}_{bound}^i(\overline{T}_{op}^{i}, \overline{T}_{lb}^{i},  \overline{T}_{ub}^{i}) & =
    \begin{cases}
    \overline{T}_{op}^i - \overline{T}_{ub}^i,  & \overline{T}_{op}^i > \overline{T}_{ub}^i \\
    \ 0,  & \overline{T}_{lb}^i \leq \overline{T}_{op}^i \leq \overline{T}_{ub}^i \\
    \overline{T}_{lb}^i - \overline{T}_{op}^i,  & \overline{T}_{op}^i < \overline{T}_{lb}^i
    \end{cases} 
    \label{equ_C5_cntrl_tracking_error_Ebound}
\end{flalign}

\vspace*{-0.25in}
\noindent where,

\vspace*{0.1in}
\noindent\begin{tabularx}{\linewidth}{lX}
        $(\cdot)^i$ & Parameter for the $i^{th}$ zone.  \\  
        $\overline{T}_{op}^{i}$ & Indoor operative temperature $(\degree C)$. \\
        $\overline{T}_{ub}^{i}$ & Upper bound on $\overline{T}_{op}^{i} (\degree C)$. \\      
        $\overline{T}_{lb}^{i}$ & Lower bound on $\overline{T}_{op}^{i} (\degree C)$. \\ 
        $\overline{T}_{set}^{i}$ & Setpoint temperature $(\degree C)$.
\end{tabularx} 

\vspace*{0.1in}

\subsection{Simulation parameters}
\label{C5_subsec_simulation parameters}


\noindent Similar to previous MPC simulation strategies discussed earlier in the thesis,  a multistep MPC feedback strategy \cite{grune2009analysis,grune2011nonlinear,Seal_2019,Seal_thesis} is used. The simulation parameters, $T_s =$ 3.75 min, $PH = $ 8 h and $CH =$ 15 min, are the simulation time step, prediction horizon and control horizon respectively. The first \textit{four} optimized control values are implemented before the MPC performs the subsequent optimization.


\subsubsection{Time-of-Use (ToU) rates and Feed-in-Tariff (FiT)}
\label{C5_subsubsec_price}

\noindent The Ontario ToU electricity rates \cite{Eprice}, are used to estimate the energy cost. The weekends have flat off-peak rates of 6.5 \textcent/kWh. During weekdays 7am - 11am and 5pm - 7pm are on-peak periods charging 13.2 \textcent/kWh, 11am - 5pm is mid-peak charging 9.4 \textcent/kWh and 7pm - 7am is off-peak period. 

The same rates are used as the Feed-in-Tariff (FiT) while considering an incentive for selling to the grid. As described in \cite{FIT_Ontario}, currently, a flat average price reduction is implemented for different KWh slabs. Here, for the simulation purposes, the FiT is chosen identical to the ToU rates.



\subsection{Description of the optimization variables}
\label{C5_subsec_Opt_Var_PV}

\noindent \Cref{C5_tab_PV_opt_var} summarizes the optimization variables and the respective ranges of each as used in the optimization. There are nine parameters which are optimized over the 8 h prediction horizon. This leads to simultaneous optimization of total $9 \times 8 \times 16 = 1152$ variables at an interval of every 15 min. 

To reduce the computational complexity and the frequent changes in the control signal, simplifying assumptions are implemented in the optimization \cite{Seal_2019,Seal_thesis}. The control variables are restricted to retain a constant value for an hour, thereby reducing the number of optimization variables is reduced by a factor of 16 (1 h = 16 time samples). As the optimization itself is repeated every 15 min, the coarseness induced by this factorization is reduced.

\vspace*{0.1in}
\noindent The power flow equations, as functions of the optimization variables (\Cref{C5_tab_PV_opt_var}), are given by the set of equations in (\ref{C5_equ_PowerFlow}).  

\vspace*{-0.1in} 
\begin{subequations}
\begin{flalign} 
P_{\!_{BB}} & = \ q_{\!_{BB}} \ =  \ q_{\!_{BB_{rated}}} \times \ F_{\!_{BB}}, & \\ 
P_{\!_{load}} & =   P_{HP_{\!_{\text{OU}}}} \ + \  P_{HP_{\!_{\text{IN,fan}}}} \ + \ P_{\!_{BB}} \ + \ P_{\!_{elec}}
\end{flalign}
\begin{flalign} 
P_{\!_{PV \rightarrow H}} & = 
	\begin{cases}
		P_{\!_{PV}} \times F_{\!_{PV \rightarrow H}}, & P_{\!_{load}} \geq P_{\!_{PV}} \times F_{\!_{PV \rightarrow H}}  \\
		P_{\!_{load}}, & P_{\!_{load}} < P_{\!_{PV}} \times F_{\!_{PV \rightarrow H}}
	\end{cases} 
		\label{C5_equ_PV2H} & \\
P_{\!_{PV \rightarrow B}} & =
	\begin{cases}
		P_{\!_{PV}} \times (1 - F_{\!_{PV \rightarrow H}}),& P_{\!_{B}} > 0; \\
		& P_{\!_{B}} \geq P_{\!_{PV}} \times (1 - F_{\!_{PV \rightarrow H}}) \\
		P_{\!_{B}},& P_{\!_{B}} > 0; \\
		& P_{\!_{B}} < P_{\!_{PV}} \times (1 - F_{\!_{PV \rightarrow H}}) \\
		0,& P_{\!_{B}} < 0
	\end{cases} 
		\label{C5_equ_PV2B} \\
P_{\!_{PV \rightarrow G}}  & = P_{\!_{PV}} - (P_{\!_{PV \rightarrow H}} + P_{\!_{PV \rightarrow B}}), \ P_{\!_{PV}} > (P_{\!_{PV \rightarrow H}} + P_{\!_{PV \rightarrow B}}) 
		\label{C5_equ_PV2G} 
\end{flalign}
\vspace*{-0.3in}
\begin{flalign} 
P_{\!_{B \rightarrow H}} & = 
	\begin{cases}
		-P_{\!_{B}} \times F_{\!_{B \rightarrow H}}, & P_{\!_{B}} < 0; \\
		& (P_{\!_{load}}  - P_{\!_{PV \rightarrow H}}) \geq -P_{\!_{B}} \times F_{\!_{B \rightarrow H}}  \\
		P_{\!_{load}} - P_{\!_{PV \rightarrow H}}, & P_{\!_{B}} < 0 ; \\ 
		& (P_{\!_{load}}  - P_{\!_{PV \rightarrow H}}) < -P_{\!_{B}} \times F_{\!_{B \rightarrow H}} \\
		0, & P_{\!_{B}} > 0
	\end{cases} 
		\label{C5_equ_B2H} & \\
F_{\!_{B \rightarrow H}} & = P_{\!_{B \rightarrow H}} \  / \ -P_{\!_{B}}, \quad \text{updating} \ F_{\!_{B \rightarrow H}} \\
P_{\!_{B \rightarrow G}} & = 
	\begin{cases}
		-P_{\!_{B}} \times (1 - F_{\!_{B \rightarrow H}}), & \ \ P_{\!_{B}} < 0 \\
		0, & \ \ P_{\!_{B}} > 0
	\end{cases} 	
	\label{C5_equ_B2G} \\
P_{\!_{G \rightarrow H}} & = P_{\!_{load}} - P_{\!_{PV \rightarrow H}} - P_{\!_{B \rightarrow H}} 
		\label{C5_equ_G2H} & \\
P_{\!_{G \rightarrow B}}  & = P_{\!_{B}} - P_{\!_{PV \rightarrow B}}, \quad \ P_{\!_{B}} > 0; \ P_{\!_{B}} > P_{\!_{PV \rightarrow B}} 
		\label{C5_equ_G2B} 
\end{flalign}
\label{C5_equ_PowerFlow}
\end{subequations}
    \noindent where,

\begin{table*}[pos=h!]
\noindent\makebox[\linewidth]{\rule{\textwidth}{0.6pt}}
\begin{flalign}
    \min_{\overline{PLR}_{\!_{IN}}, \ \overline{F}_{\!_{BB}}, \ \overline{P}_{\!_{B}}, \ \overline{F}_{\!_{PV \rightarrow H}}, \ \overline{F}_{\!_{B \rightarrow H}}} \ & \Big[\underbrace{w_1. \overline{c}^T_{buy} . (\overline{P}_{\!_{G \rightarrow H}} + \overline{P}_{\!_{G \rightarrow B}}) \ - \ w_2.\overline{c}^T_{sell} . (\overline{P}_{\!_{PV \rightarrow G}} + \overline{P}_{\!_{B \rightarrow G}})}_{J_{\text{cost}}} \ + \ w_3. \sum_{k = 2}^{L} \Delta SOC_{\!_B}(k) \ \nonumber & \\ 
\  \ & \hspace*{1.5in} + \underbrace{\sum_{i = 1}^{z} w^i_4. \| \ \overline{e}^i_{bound}(\overline{T}^i_{op}, \overline{T}_{lb}, \overline{T}_{ub}) \ \|_2 + \sum_{i = 1}^{z} w^i_5. \sum_{k = 1}^{L}  \overline{e}^{i,k}_{set}(\overline{T}^{i,k}_{op}, \overline{T}^k_{set})}_{J_{\text{comfort}}} \ \Big]
    \label{C5_eqn_cost_function_PV}
\end{flalign}
%
%
\vspace*{-.2in} \flushleft Where, \vspace*{.1in} \\ 
\noindent\begin{tabularx}{\linewidth}{XXX}
	    $\overline{PLR}_{\!_{IN}}, \ \overline{F}_{\!_{BB}} \in \mathbb{R}^{L \times z}$; 
	    & \ $\overline{P}_{\!_{B}}, \ \overline{F}_{\!_{PV \rightarrow H}}, \ \overline{F}_{\!_{B \rightarrow H}}, \ \overline{c}_{buy}, \ \overline{c}_{sell} \in \mathbb{R}^L$;  
	    & $\overline{P}_{\!_{G \rightarrow H}}, \ \overline{P}_{\!_{G \rightarrow B}}, \ \overline{P}_{\!_{PV \rightarrow G}}, \ \overline{P}_{\!_{B \rightarrow G}} \in \mathbb{R}^L$; \\
	    $i  = 1, \dots, z; \quad  k  = 1, \dots, L$; 
	    & \ $\overline{e}^i_{bound}, \ \overline{T}^i_{op}, \ \overline{T}_{lb}, \ \overline{T}_{ub} \in \mathbb{R}^L$;   
	    & $\overline{e}^{i,j}_{set}, \ \overline{T}^{i,j}_{op}, \ \overline{T}^j_{set}, \ w_{1, 2, 3}, \ w^i_{4, 5}  \in \mathbb{R}$ \\
	    $\Delta SOC_{\!_B}(k) = \big| SOC_{\!_B}(k - 1) - SOC_{\!_B}(k) \big|$; 
	    & $\ SOC_{\!_{B_{min}}} \ \leq \ SOC_{\!_B}(k) \ \leq \ SOC_{\!_{B_{max}}}$
\vspace*{0.1in}
    \end{tabularx}
\noindent\makebox[\linewidth]{\rule{\textwidth}{0.6pt}}
 \end{table*}
 
\vspace*{0.1in}
\noindent\begin{tabularx}{\linewidth}{lX}
    $P_{\!_{BB}}$ & Power consumption by the baseboards (kW).  \\
    $q_{\!_{BB}}$ & Heating capacity of the baseboards (kW). \\
    $P_{\!_{load}}$ & Load demand including HVAC (kW).  \\
    $P_{HP_{\!_{\text{OU}}}}$ & Effective power consumption (kW) by the outdoor unit of the multi-split heat pump system, as given in (\ref{C5_equ_PHP_OU}) (kW). \\ 
    $P_{HP_{\!_{\text{IN,fan}}}}$ & Total fan power for indoor units of the multi-split system = 30 W, for each zone. \\
    $P_{\!_{elec}}$ & Power consumption by electrical appliances as indicated in \Cref{C5_fig_Bungalow_electrical_load}  (kW). 
\end{tabularx}
\noindent\begin{tabularx}{\linewidth}{lX}
    $P_{\!_{PV}}$ & Power generated by the PV (kW). \\
    $P_{\!_{B}}$ & Power delivered (or consumed) during discharging (or charging) (kW). \\
    $P_{\!_{PV \rightarrow H}}$ & Power delivered from PV to house (kW).  \\
    $P_{\!_{PV \rightarrow B}}$ & Power delivered from PV to home battery (kW). \\
    $P_{\!_{PV \rightarrow G}}$ & Power delivered from PV to grid (kW). \\
    $P_{\!_{B \rightarrow H}}$ & Power delivered from home battery to house (kW).  \\
    $P_{\!_{B \rightarrow G}}$ & Power delivered from home battery to grid (kW). \\
    $P_{\!_{G \rightarrow H}}$ & Power delivered from grid to house (kW). \\
    $P_{\!_{G \rightarrow B}}$ & Power delivered from grid to home battery (kW). 
\end{tabularx}

\vspace*{0.1in}    
\noindent $P_{\!_B}$ is considered to be positive during charging, negative during discharging and zero for the neutral state. 95\% home battery efficiency is considered to calculate the battery operations. The state of charge of the home battery ($SOC_{\!_B}$) is maintained between 10\% - 90\% at every time step.

 
\section{Case study scenarios and objective function}
\label{C5_sec_PV_objectiveFunction_caseStudy}

\noindent The primary objective of the MPC is to reduce energy consumption from the grid. The goal is to supply sufficient energy, to the heating unit of the house, in order to maintain the indoor temperature within prescribed limits along with satisfying other daily load demands. The MPC objective function is given in (\ref{C5_eqn_cost_function_PV}). Here,

\vspace*{0.1in}
\noindent\begin{tabularx}{\linewidth}{lX}
    $z$ & No. of zones in the house. \\
    $L$ & Length of $PH$. \\
    $\overline{c}_{buy}$ & Time-of-Use rates (\textcent/kWh). \\
    $\overline{c}_{sell}$ & Feed-In Tariff (\textcent/kWh). \\
    $w_{1,2,3,4,5}$ & Weights on the different penalty. \\
    $\Delta SOC_{\!_{B}}$ & Change in the state of charge of the battery at each time step (kWh). 
\end{tabularx}

\vspace*{0.1in}      
\noindent The cost term, $J_{\text{cost}}$ gives the total energy cost corresponding to the net energy consumption from the grid over a given prediction horizon $PH$. The grid supplies the load demand from the house and charges the battery as needed. The timing and amount of power delivered from the grid is controlled by the MPC, by optimizing the variables $F_{\!_{PV \rightarrow H}}$ and $F_{\!_{B \rightarrow H}}$ as given in (\ref{C5_equ_PowerFlow}). The comfort term, $J_{\text{comfort}}$ in (\ref{C5_eqn_cost_function_PV}) includes the $l_2$-norm of the deviation of $T_{op}$ from the upper and lower temperature bounds and the absolute error of deviation from the setpoint as defined by \cref{equ_C5_cntrl_tracking_error_Eset,equ_C5_cntrl_tracking_error_Ebound}. 

The weights on the cost and comfort terms ($w_1, \dots ,w_5$), for respective zones, are used to scale the components of the objective function having different units and thus adjust the performance of the MPC controller. These are determined by trial and error to achieve a better trade-off between energy cost and comfort. These weights are tunable parameters and the choices are based on user preference. Since this is a multivariable control optimization and a single objective function is used to optimize the MPC performance for the three zones simultaneously, the weights set priority on the goals to achieve the desired performance. The weights are specified as follows: 

\vspace*{0.1in}
\noindent\begin{tabularx}{\linewidth}{lX}
$w_1$ & Weight on $c_{buy}$. \\
$w_2$ & Weight on $c_{sell}$. \\
$w_3$ & Weight on $\Delta SOC_{\!_{B}}$. \\
$w_4^i$ & Weights on the deviation from $T_{bound}$ for $i^{th}$ zone. 
\end{tabularx}
\noindent\begin{tabularx}{\linewidth}{lX}
$w_5^i$ & Weights on the deviation from $T_{set}$ for $i^{th}$ zone.
\end{tabularx}

\subsection{Base case scenario}
\label{C5_subsec_PV_BaseCase}

\begin{figure}[pos=h!]
\vspace*{-0.3in}
\hspace*{-0.3in}
    \includegraphics[scale = 0.8]{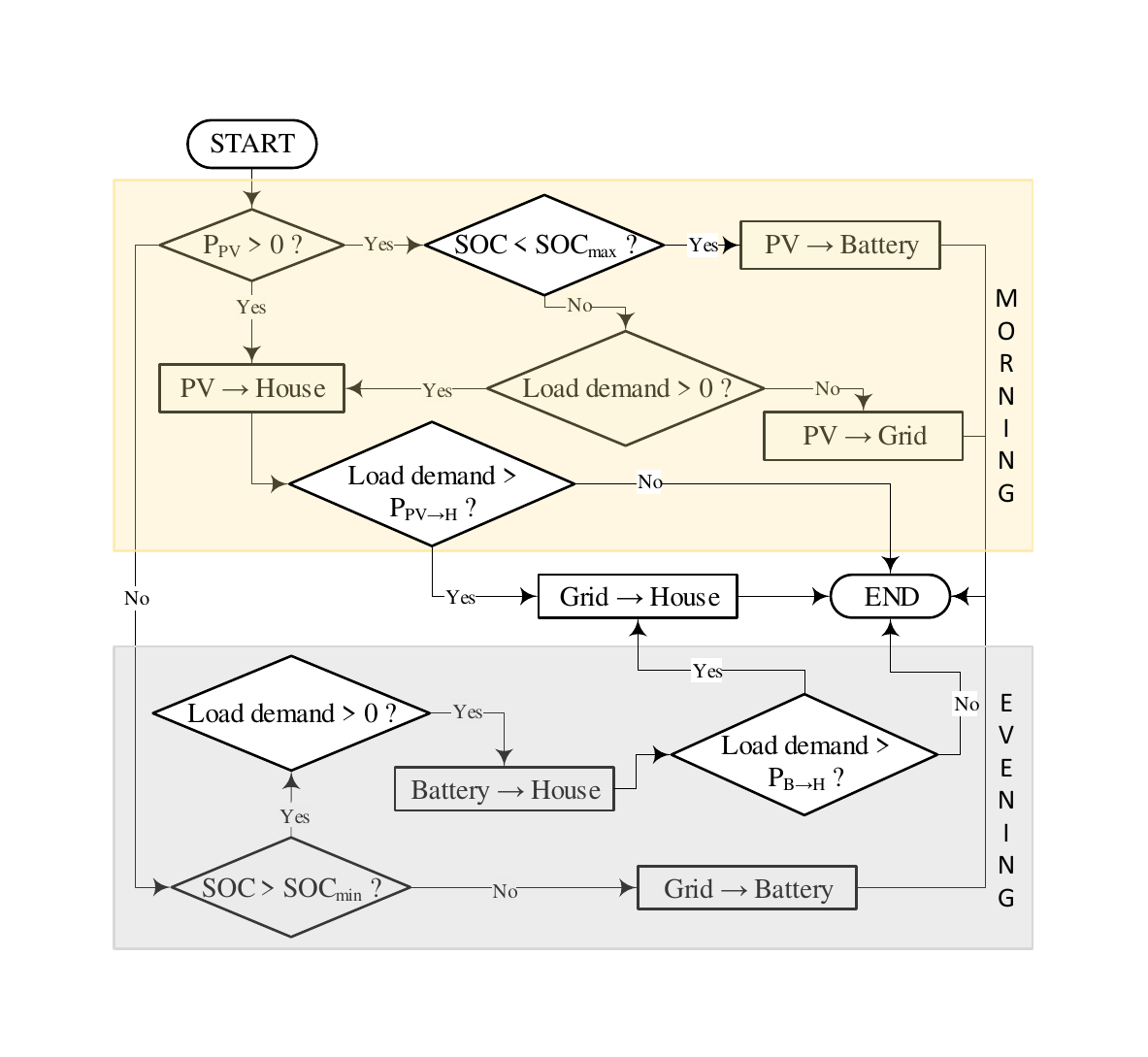}
    \vspace*{-0.4in}
    \caption{Rule based energy flow diagram for the base case scenario with PV and home battery}
    \label{C5_fig_base_PV_Bat}
\vspace*{-0.15in}
\end{figure}

\noindent An MPC controlled comfort management system is considered in the base case scenario along with a rule based energy management strategy described in the flowchart in \Cref{C5_fig_base_PV_Bat}. The assumptions for the controller strategy for the base case scenario are as follows:

\begin{itemize}[noitemsep]

\item The power inputs to the heating units, i.e., $PLR_{\!_{IN}}$, $F_{\!_{BB}}$ and $P_{\!_{B}}$, specifically the battery discharge power $PB_{\text{discharge}}$ only are controlled by the MPC.

\item The power consumptions by the HP and BB heating units are optimized based on the variable ToU rates.



\item A minimum 50\% battery SOC is always maintained for emergency usage, i.e., $SOC_{\!_{B_{min}}} = 50\%$. 


\item Grid energy flow is bidirectional. Excess PV generated electricity can be fed back to the grid but without any incentive.

\end{itemize}

\noindent The flowchart in \Cref{C5_fig_base_PV_Bat} shows the energy flow diagram to and from PV, home battery, house and grid, for a single time step during the optimization in the base case scenario. The energy flow is primarily rule based. 


\subsubsection{Case I : Bidirectional grid connection with an incentive for selling power to the grid}
\label{C5_subsubsec_MPC_Sell}

\noindent This scenario considers bidirectional communication with the power grid. A two-way grid connection is available. A Feed-in-Tariff (FiT) same as the ToU rates (\Cref{C5_subsubsec_price}) is used. For this particular scenario, the weights on the objective functions defined in (\ref{C5_eqn_cost_function_PV}) are modified as follows: 

\vspace*{0.1in}
\noindent\begin{tabularx}{\linewidth}{llX}
 $w_1$ & = &1.5; \\ 
 $w_2$ & = & 1; \\
 $w_3$ & = & 5; \\
 $w_4^{1,2,3}$ & = & 3.2, same for all zones. \\
 $w^{1,2,3}_5$ & = & \{0.08, \ 0.05, \ 0.08\}, for the primary and secondary bedroom, and the living room respectively.
\end{tabularx}



\section{Results and discussion}
\label{C5_sec_result}

\noindent The performances of the MPC, for the above mentioned scenarios are analyzed for a 7-day simulation period. The simulation is run for 8 consecutive days during the first week of January and the first day is considered as a warm-up day. The TRNSYS-MATLAB co-simulation used for this study takes approximately 3 days in real-time to simulate 8 simulation days on an Intel Core-i7 processor with 8 GB RAM. Though computationally time-expensive, the controller is real-time implementable. But the extended simulation time restricts the choice of a longer simulation interval. In the following, the three case study scenarios are compared based on three performance criteria :

\begin{itemize}[noitemsep]
	\item \textbf{Comfort performance:} The Time-in-Target performance of the MPC in maintaining the operative temperature of the three zones within the comfort bounds, as mentioned in \Cref{C5_subsec_setpoint_heating}.
	
	\item \textbf{Heating energy distribution:} The distribution of the heating energy between the multi-spit heat pump (HP) unit and the electric baseboard (BB) system.
	
	\item \textbf{Use of renewable solar energy and battery storage unit:} This considers the performance of the controller in terms of utilization of the solar energy generated by the PV to supply the load demand of the house and to charge the battery. Also, the charging and discharging cycle of the battery and use of the storage unit in distributing the peak demand to avoid the on-peak electricity supply from the grid. 
		
\end{itemize}

 
\subsection{Comfort performance}
\label{C5_subsec_result_comfort}

\begin{figure*}[pos=h!]
    \centering
    \includegraphics[scale = 0.8]{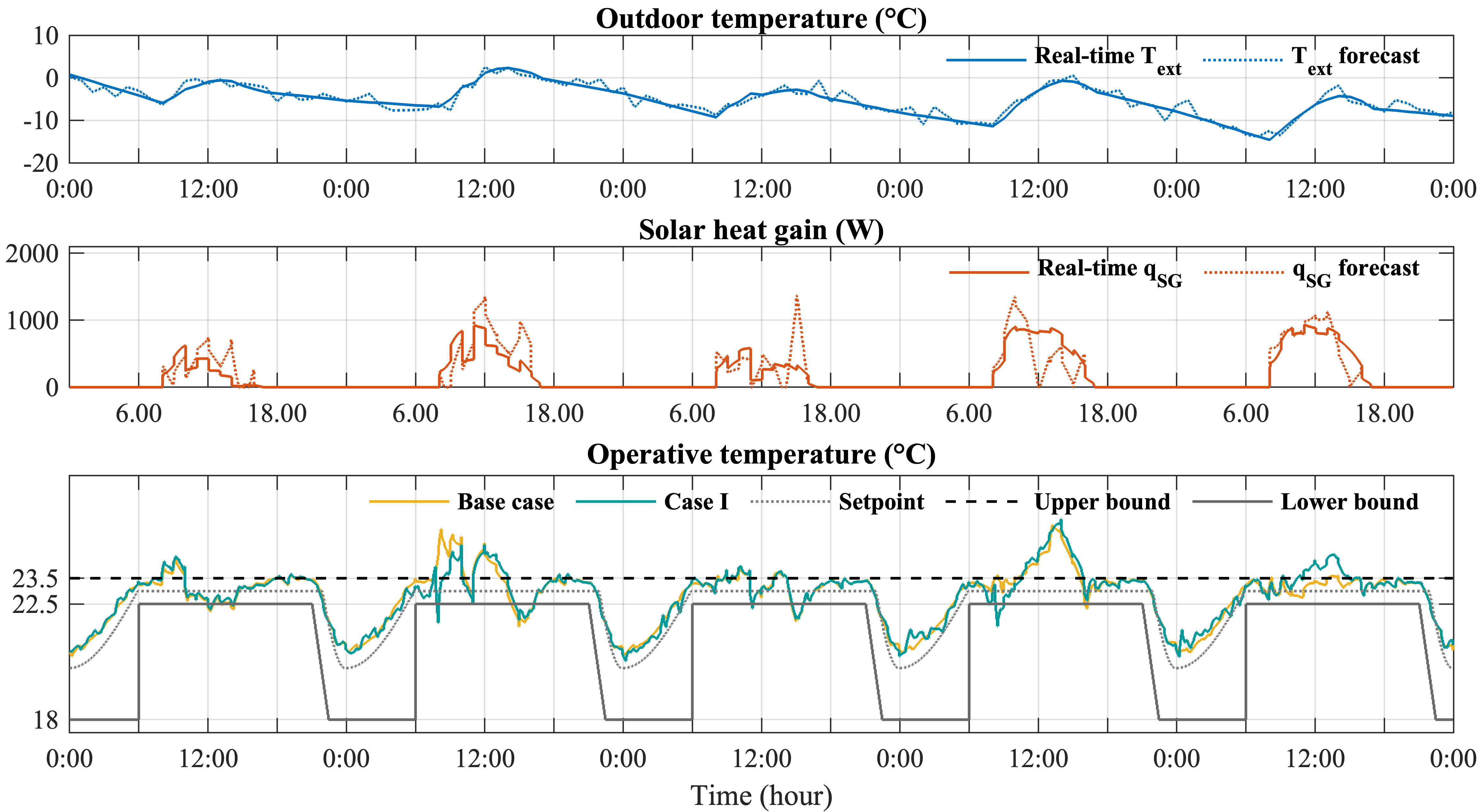}
    \caption{Operative temperature ($T_{op}$) profile for the primary bedroom (B1) for the three simulation scenarios along with the real-time and forecast dataset for outdoor temperature, $T_{ext}$ and solar heat gain $q_{\!_{SG}}$, for 5 consecutive days as an example.}
    \label{C5_fig_comfort_PV_B1}
\end{figure*}

\noindent \Cref{C5_fig_comfort_PV_B1} shows the operative temperature plots for the base case and Case I. Five consecutive days of simulation results are shown in the figure, as an example, for the primary bedroom (B1). The forecast and real-time outdoor temperature and respective solar heat gain for B1 are also indicated. 

\begin{figure}[pos=h!]
    \centering
    \includegraphics[scale = 0.65]{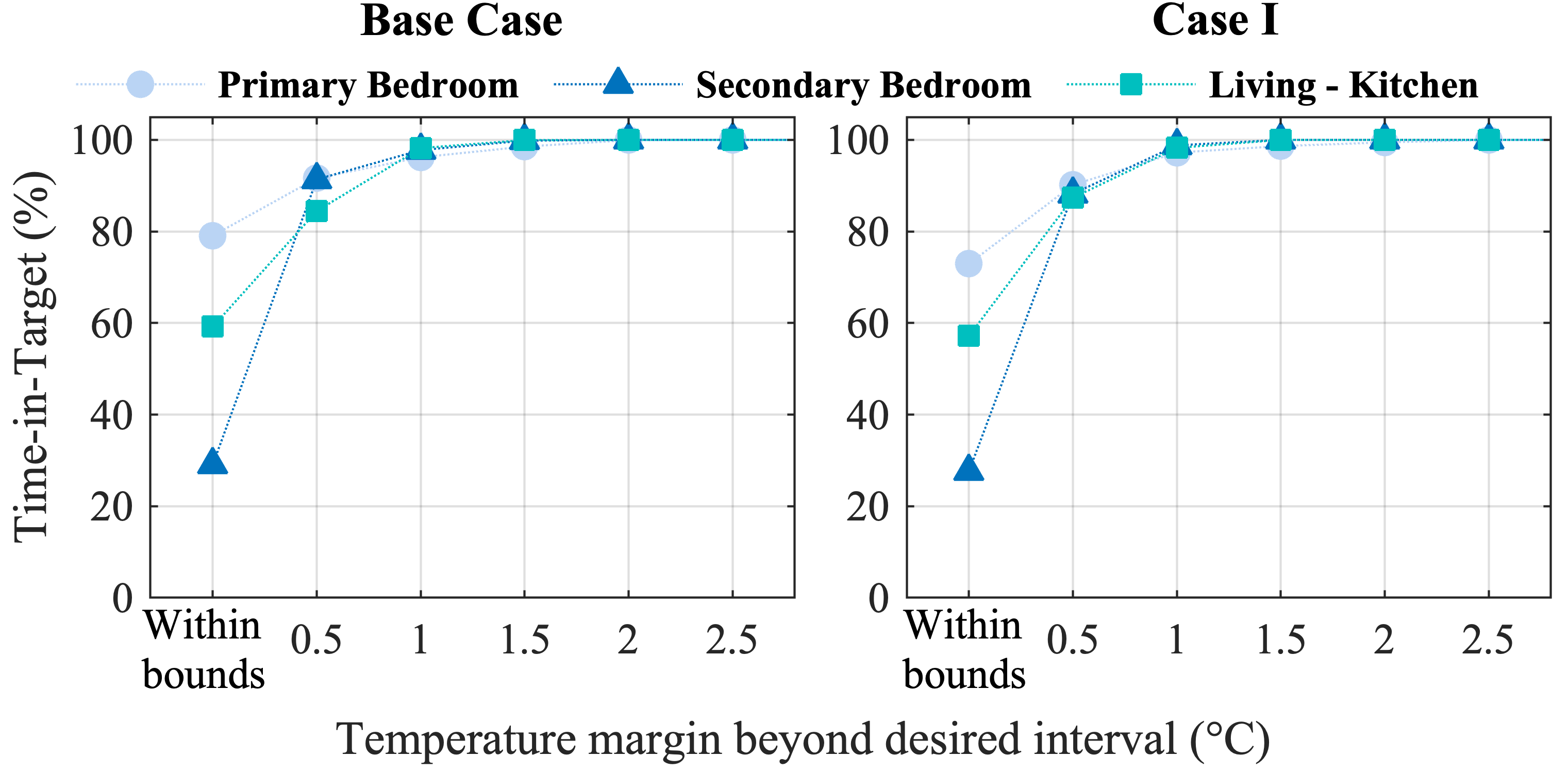}
    \caption{Cumulative percentage showing Time-in-Target performance of the MPC, to maintain the operative temperatures, $T_{op}$ within the comfort bounds, for the 7-day simulation period.}
    \vspace*{-0.15in}
    {\justify {\footnotesize Here, the horizontal axis indicates the additional margin from either the upper or lower temperature bound defining the new target range. For example, the $0.5 \degree C$ margin covers the range : $T_{lb} - 0.5 \leq T_{op} \leq T_{ub} + 0.5$ \par}}
    \label{C5_fig_comfort_hist_PV_3zone}
\end{figure}

The MPCs perform somewhat similarly, in maintaining the $T_{op}$ within the specified limit, in both the cases. This is also reflected by the Time-in-Target performance in \Cref{C5_fig_comfort_hist_PV_3zone}, where the histogram of the deviations of $T_{op}$, in each case, is indicated for the entire 7-day simulation period.

Certain observations regarding the effects of the forecast uncertainty can be summarized from \Cref{C5_fig_comfort_PV_B1}. The temperature plot on the $4^{th}$ day shows a deviation of close to $2 \degree C$ resulting from the significant dissimilarity in the forecast prediction for the solar heat gain. The MPC heats the zone considering a lower heat gain, which eventually results in overheating due to the solar radiation. Since no forced cooling is available, it takes a while before the $T_{op}$ lower to reach the desired limit. Similarly, insufficient heating also results from over-prediction of the solar heat gain in the zone, as can be seen on the $2^{nd}$ or the $3^{rd}$ day.

\noindent The Time-in-Target shown in \Cref{C5_fig_comfort_hist_PV_3zone}, is based on the deviation from the upper and lower temperature bounds only. The $e_{set}$ defined in (\ref{equ_C5_cntrl_tracking_error_Eset}) with reference to the $T_{set}$ in \Cref{fig_C5_setpoint}, is used in the objective function to induce a penalty on the overheating effect, but no related penalty is considered in calculating the Time-in-Target performance. 

Above 85\% of the time, the $T_{op}$ in each zone, is maintained within a $\pm 0.5 \degree C$ deviation margin, with a maximum deviation of close to $\pm 1 \degree C$. The secondary bedroom (B2), because of its North-East orientation (\Cref{fig_C5_Bungalow}), has less solar heat gain than the primary bedroom (B1) and the living-kitchen (LK). Hence, the MPC maintains the $T_{op}$ of zone B2, closer to the upper temperature bound in order to achieve a trade-off between the heating energy cost and the comfort performance. This leads to respectively 30\% and 50\% reduction in the Time-in-Target performance, as compared to the zones B1 and LK. However, the MPC performance in zone B2 matches with the other two zones while comparing with respect to the range of $\pm 0.5 \degree C$ deviation margin. 



\subsection{Heating energy distribution}
\label{C5_subsec_result_heating_energy}


\begin{figure}[pos=h!]
    \centering
    \includegraphics[scale = 0.7]{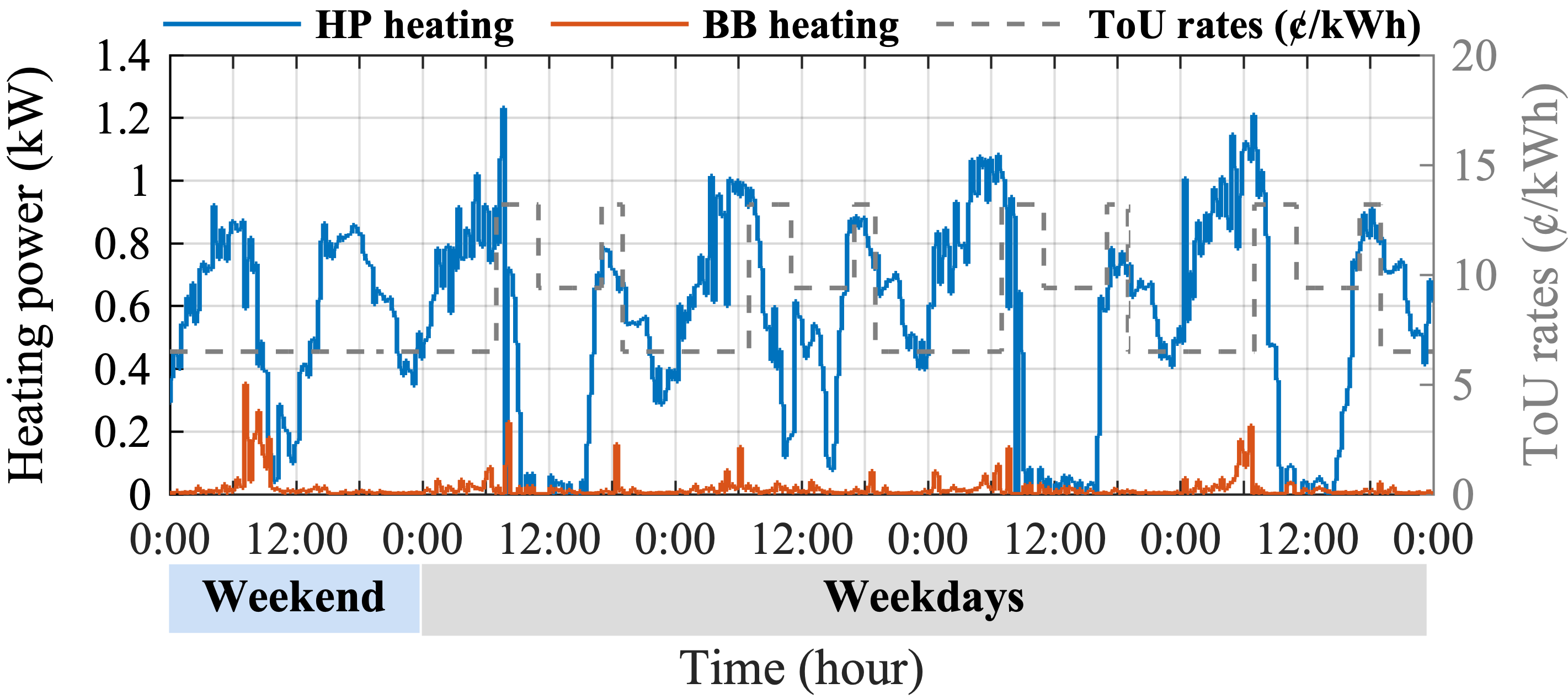}
    \caption{Example of HP and BB heating pattern: heating for zone B1 in Case I for a 5-day simulation period.}
    \label{C5_fig_Pow_HPBB_PV_B1}
\end{figure}

\noindent The distribution of the heating energy between the HP and the BB heating units are also similar for the two scenarios. \Cref{C5_fig_Pow_HPBB_PV_B1} shows a sample heating pattern in the Case I scenario for zone B1. The HP is used as the primary heating system since the average COP of the HP is 3.26 as compared to unity in case of the BB. The BB is used as a secondary system to provide any additional heating if needed during the day. HP preheats the zone to take advantage of the off-peak ToU rates, specially during the night. 

\begin{figure}[pos=h!]
    \centering
    \includegraphics[scale = 0.7]{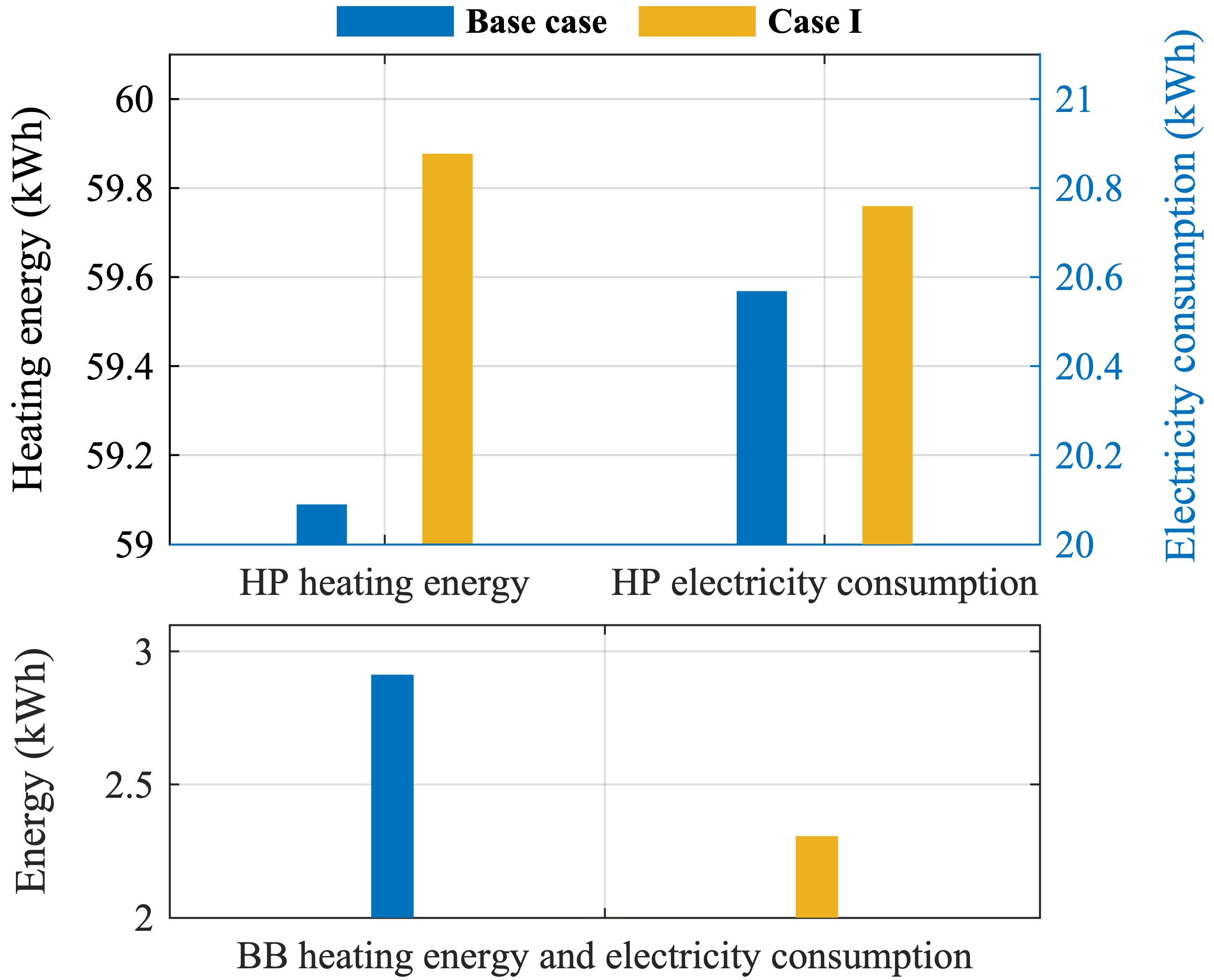}
    \caption{Daily average HP and BB heating energy delivered to the house and corresponding electricity consumption for the two case study scenarios during the 7-day simulation period.}
    \label{C5_fig_Energy_HPBB_PV}
\vspace*{-0.1in}
\end{figure}

\noindent The HP net electricity consumptions are shown in \Cref{C5_fig_Energy_HPBB_PV}. Since 100\% efficiency of the BB is assumed in general, the heat delivered by the BB is identical to its electricity consumption. The differences in the amounts of heating energy usage by the house in the two cases are minimal. The base case uses more BB heating. Based on the average heating energy delivered by the HP and BB,  the average daily per unit cost of electricity for BB usage is approximately 20\% more than that of HP usage for both the cases. 

 
\subsection{Use of renewable solar energy and battery storage unit}
\label{C5_subsec_result_energy_flow}

\begin{figure}[pos=h!]
    \centering
    \includegraphics[scale = 0.75]{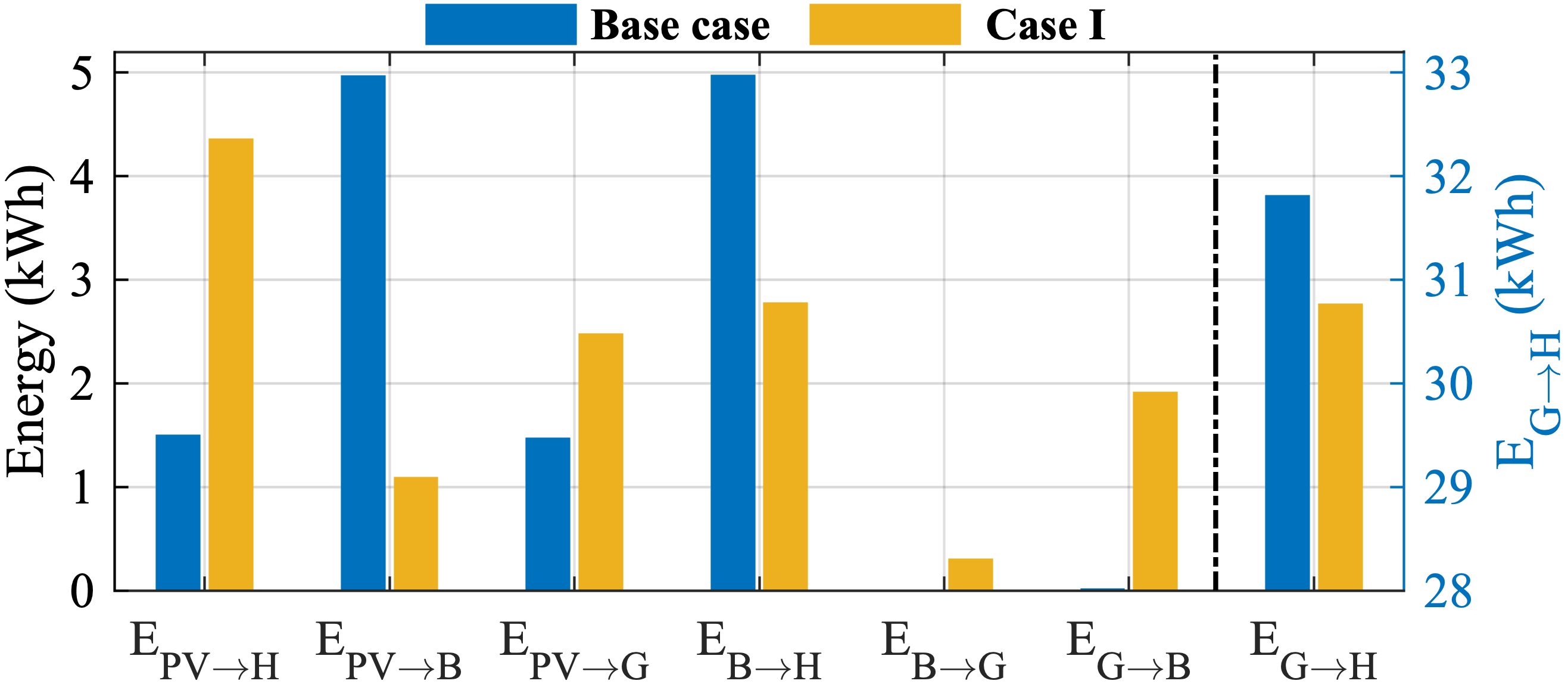}
    \caption{Daily average energy flow from PV, home battery and the grid during the 7-day simulation period.}
    \label{C5_fig_Energy_flow_PV}
\end{figure}

\noindent The daily average energy flows from different sources to sinks are shown in \Cref{C5_fig_Energy_flow_PV}. Here, the labels on the horizontal axis are given by,

\noindent\begin{tabularx}{\linewidth}{lX}
    $E_{\!_{PV \rightarrow H}}$ & Energy delivered from PV to house (kWh). \\
    $E_{\!_{PV \rightarrow B}}$ & Energy delivered from PV to home battery (kWh). \\
    $E_{\!_{PV \rightarrow G}}$ & Energy delivered from PV to grid (kWh). \\
    $E_{\!_{B \rightarrow H}}$ & Energy delivered from home battery to house (kWh). \\
    $E_{\!_{B \rightarrow G}}$ & Energy delivered from home battery to grid (kWh). \\
    $E_{\!_{G \rightarrow H}}$ & Energy delivered from grid to house (kWh). \\
    $E_{\!_{G \rightarrow B}}$ & Energy delivered from grid to home battery (kWh). 
\end{tabularx}

\vspace*{0.1in}
\noindent The rule based simulation scenario presented in the base case (\Cref{C5_subsec_PV_BaseCase}), drives the PV generated energy to charge the battery in the morning as a priority. The battery is mostly used to deliver energy to the house at night and the MPC controls the amount of discharge. 

\begin{figure}[pos=h!]
    \centering
    \includegraphics[scale = 0.76]{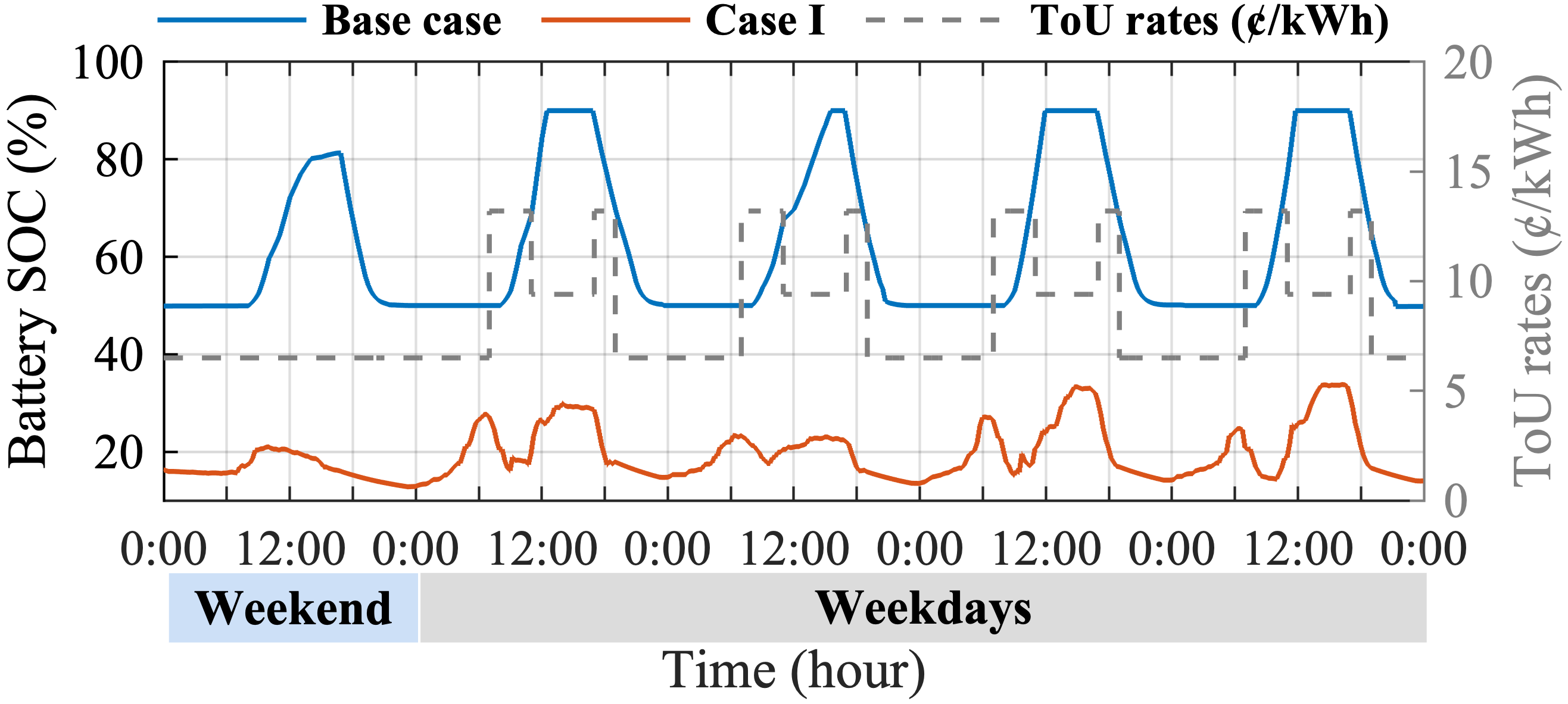}
    \caption{State of charge (SOC) of the battery for 5 consecutive days, corresponding to \Cref{C5_fig_comfort_PV_B1}}
    \label{C5_fig_SOC_base_case1}
\end{figure}

In comparison to the base case, the centralized MPC strategy in Case I, utilizes a larger share of the PV generated electricity to supply the instantaneous load demand of the house. $E_{\!_{PV \rightarrow H}}$ for the base is close to 66\% less than Case I. Whereas, the rule based control in the base case utilizes the PV generated power mostly to charge the battery. The $E_{\!_{PV \rightarrow B}}$ in Case I is 78\% less than the base case. The variation in $SOC_{\!_{B}}$ is plotted in \Cref{C5_fig_SOC_base_case1}.

$SOC_{\!_{B}}$ is maintained within 50\% to 90\% for the base case. The battery charging is more flexible and may vary between 10\% - 90\% for Case I. 50\% SOC level was considered as the initial condition. Here, the MPC charging strategy follows a general pattern of `two-phase' charging. Initially, utilizing the off-peak prices at the start of the day, the battery is charged to supply the energy demand during the day-time on-peak period. In the second phase, the battery is charged using solar energy and supplies the demand during the second on-peak period in the evening. The first day, indicated in the \Cref{C5_fig_SOC_base_case1}, is a weekend and the ToU rate off-peak throughout the day. As indicated in \Cref{C5_fig_comfort_PV_B1}, the day is also cloudy. The MPC decides to minimally charge the battery to maintain the 10\% SOC. Overall, to achieve the objective of minimizing the energy cost, the MPC tends to utilize approximately 10\% - 50\% of the battery capacity and charges the battery only sufficiently to reduce the grid energy during the peak ToU periods. 

\begin{table}[pos=h!]
    \caption{Daily average energy cost for 7-day simulation period}
     \label{C5_tab_Bungalow_Energy_Cost}
{\setlength\extrarowheight{0.1pt}
    {\scriptsize\addtolength{\tabcolsep}{-6pt}
    \begin{tabular*}{\tblwidth}{@{} CCCCCCC@{} } 
    \toprule 
     \multicolumn{1}{ l }{ } &
     \multicolumn{1}{ p{0.5cm} }{} &
     \multicolumn{1}{ l }{\textbf{\  Base case \ }} &
     \multicolumn{1}{ p{0.5cm} }{} &
     \multicolumn{1}{ l }{\textbf{\ Case I \ }} \\ \midrule 
     \multicolumn{1}{ l }{\ Average daily electricity buying cost (C\$) \ } &
     \multicolumn{1}{ p{0.5cm} }{} & 
     \multicolumn{1}{ l }{\ 2.52 \ } &
     \multicolumn{1}{ p{0.5cm} }{} &     
     \multicolumn{1}{ l }{\ 2.45 \ } \\   
     \multicolumn{1}{ l }{\ Revenue from selling electricity (C\$) \ } &
     \multicolumn{1}{ p{0.5cm} }{} &
     \multicolumn{1}{ l }{\ - \ } &
     \multicolumn{1}{ p{0.5cm} }{} &  
     \multicolumn{1}{ l }{\ 0.27 \ } \\  \midrule 
     \multicolumn{1}{ l }{\ Net daily average expense (C\$) \ } &
     \multicolumn{1}{ p{0.5cm} }{} &
     \multicolumn{1}{ l }{\ 2.52 \ } &
     \multicolumn{1}{ p{0.5cm} }{} &     
     \multicolumn{1}{ l }{\ 2.18 \ }  \\ \bottomrule 
     \end{tabular*}}}
\end{table}

In general, the usage of grid energy is more in Case I, where the effect of the extra energy cost for buying energy from the grid is balanced by selling energy back to the grid. The average daily electricity costs are listed in \Cref{C5_tab_Bungalow_Energy_Cost}. The minimum is achieved in Case I, where a FiT is implemented for selling energy to the grid.

In Case I, the simultaneous comfort and BEMS MPC optimization achieve a 13.5\% reduction in the energy cost as compared to the base case. A simple case of MPC optimization for comfort management (without any BEMS control) is also simulated in TRNSYS as a reference to estimate the benefit of the centralized MPC controlled comfort and energy management system. Neither PV generation nor battery storage is considered in this case. The average daily cost achieved here is 3.16 C\$, which is close to 31\% more than Case I.


\section{Conclusion}
\label{C5_sec_conclusion}

In an effort to design a centralized system to control different components of the HVAC and BEMS systems simultaneously, a novel predictive control strategy is proposed in this article. By optimizing a single multi-objective cost function, the MPC successfully maintains the operative temperature within the predefined comfort bounds and achieves a 31\% reduction in the daily average energy cost as compared to the grid-only building energy supply. The installation cost for the PV and the battery are not considered for these case studies. 

The multi-split air-to-air heat pump has a fast response time and maintains an average COP of 3.26 during the simulation period. Thus, the heat pump, along with the baseboard as a secondary heating unit builds an efficient and heating unit with low operative cost. The MPC controlled comfort and energy management system uses the PV generated solar energy to supply close to 10\% of the total heating demand of the house, thereby saving approximately 198 gCO$_2$ emission daily, considering 32 gCO$_2$/kWh GHG emission factor for Ontario \cite{Co2_factor}. The reduction in the daily energy cost with respect to the grid based energy supply system indicates long term benefits in terms of cost minimization, leading to a cleaner environment. 

In this paper, the objective function optimizes energy cost and comfort. A modification in the objective function by incorporating a penalty on the GHG emission may achieve a further reduction in the CO$_2$ emission. Specially in case of a mixed source energy generation scenario as in Ontario \cite{Canada_energy_ON}, a penalty on the GHG emission may play an effective contribution in achieving a reduction in the carbon footprint. However, for a greener grid scenario as in Quebec where more than 99\% \cite{Canada_energy_QC} of the total electricity generation is based on renewables, a cost-based objective function might be better appreciated. In summary, the MPC optimization method presented in this paper can be customized to incorporate various related control objectives to achieve user specific goals.



\section*{Acknowledgment}

The authors would like to thank National Science and Engineering Research Council (NSERC) and Natural Resources Canada (NRCan) for funding this project. We convey our sincere thanks to Mr. Justin Tamasauskas from CanmetENERGY, NRCan, Varennes, for his contributions in developing the TRNSYS building model. 

%
%
\printcredits

\bibliographystyle{cas-model2-names}
\balance
\bibliography{Paper_Chap5_ENERGY}

\begin{thebibliography}{28}
\expandafter\ifx\csname natexlab\endcsname\relax\def\natexlab#1{#1}\fi
\providecommand{\url}[1]{\texttt{#1}}
\providecommand{\href}[2]{#2}
\providecommand{\path}[1]{#1}
\providecommand{\DOIprefix}{doi:}
\providecommand{\ArXivprefix}{arXiv:}
\providecommand{\URLprefix}{URL: }
\providecommand{\Pubmedprefix}{pmid:}
\providecommand{\doi}[1]{\href{http://dx.doi.org/#1}{\path{#1}}}
\providecommand{\Pubmed}[1]{\href{pmid:#1}{\path{#1}}}
\providecommand{\bibinfo}[2]{#2}
\ifx\xfnm\relax \def\xfnm[#1]{\unskip,\space#1}\fi
\bibitem[{Beltrami et~al.(2015)Beltrami, Matharoo and
  Smerdon}]{beltrami2015ground}
\bibinfo{author}{Beltrami, H.}, \bibinfo{author}{Matharoo, G.S.},
  \bibinfo{author}{Smerdon, J.E.}, \bibinfo{year}{2015}.
\newblock \bibinfo{title}{Ground surface temperature and continental heat gain:
  uncertainties from underground}.
\newblock \bibinfo{journal}{Environmental Research Letters}
  \bibinfo{volume}{10}, \bibinfo{pages}{014009}.
\bibitem[{Candanedo et~al.(July 9-12, 2018)Candanedo, Hardy, Saloux, Platon,
  Raissi-Dehkordi and C{\^o}t{\'e}}]{candanedo2018preliminary}
\bibinfo{author}{Candanedo, J.A.}, \bibinfo{author}{Hardy, J.M.},
  \bibinfo{author}{Saloux, {\'E}.}, \bibinfo{author}{Platon, R.},
  \bibinfo{author}{Raissi-Dehkordi, V.}, \bibinfo{author}{C{\^o}t{\'e}, A.},
  \bibinfo{year}{July 9-12, 2018}.
\newblock \bibinfo{title}{{Preliminary Assessment of a Weather Forecast Tool
  for Building Operation}}, in: \bibinfo{booktitle}{{5th International High
  Performance Buildings Conference, Purdue, USA}}.
\bibitem[{Chen et~al.(2013)Chen, Wang, Heo and Kishore}]{Chen6575202_EMS}
\bibinfo{author}{Chen, C.}, \bibinfo{author}{Wang, J.}, \bibinfo{author}{Heo,
  Y.}, \bibinfo{author}{Kishore, S.}, \bibinfo{year}{2013}.
\newblock \bibinfo{title}{{MPC-Based Appliance Scheduling for Residential
  Building Energy Management Controller}}.
\newblock \bibinfo{journal}{IEEE Transactions on Smart Grid}
  \bibinfo{volume}{4}, \bibinfo{pages}{1401--1410}.
\bibitem[{Duffie and Beckman(2013)}]{duffie2013solar}
\bibinfo{author}{Duffie, J.A.}, \bibinfo{author}{Beckman, W.A.},
  \bibinfo{year}{2013}.
\newblock \bibinfo{title}{Solar engineering of thermal processes}.
\newblock \bibinfo{publisher}{New York: John Wiley \& Sons}.
\bibitem[{{Energy Sage}()}]{PV_panel}
\bibinfo{author}{{Energy Sage}}, .
\newblock \bibinfo{title}{{Size and weight of solar panels}}.
\newblock
  \bibinfo{howpublished}{\url{https://news.energysage.com/average-solar-panel-size-weight/}}.
\newblock \bibinfo{note}{[Accessed on 16-12-2018]}.
\bibitem[{Filliard et~al.(July 27-30, 2009)Filliard, Guiavarch and
  Peuportier}]{filliard2009performance}
\bibinfo{author}{Filliard, B.}, \bibinfo{author}{Guiavarch, A.},
  \bibinfo{author}{Peuportier, B.}, \bibinfo{year}{July 27-30, 2009}.
\newblock \bibinfo{title}{Performance evaluation of an air-to-air heat pump
  coupled with temperate air-sources integrated into a dwelling}, in:
  \bibinfo{booktitle}{Eleventh International IBPSA Conference Glasgow,
  Scotland}, pp. \bibinfo{pages}{2266--2273}.
\bibitem[{{Government of Canada}(a)}]{Canada_energy_ON}
\bibinfo{author}{{Government of Canada}}, a.
\newblock \bibinfo{title}{{Canada Energy Regulator: Provincial and Territorial
  Energy Profiles - Ontario}}.
\newblock
  \bibinfo{howpublished}{\url{https://www.cer-rec.gc.ca/nrg/ntgrtd/mrkt/nrgsstmprfls/on-eng.html}}.
\newblock \bibinfo{note}{[Accessed on 21-11-2019]}.
\bibitem[{{Government of Canada}(b)}]{Canada_energy_QC}
\bibinfo{author}{{Government of Canada}}, b.
\newblock \bibinfo{title}{{Canada Energy Regulator: Provincial and Territorial
  Energy Profiles - Quebec}}.
\newblock
  \bibinfo{howpublished}{\url{https://www.cer-rec.gc.ca/nrg/ntgrtd/mrkt/nrgsstmprfls/qc-eng.html}}.
\newblock \bibinfo{note}{[Accessed on 21-11-2019]}.
\bibitem[{Gr{\"u}ne(2009)}]{grune2009analysis}
\bibinfo{author}{Gr{\"u}ne, L.}, \bibinfo{year}{2009}.
\newblock \bibinfo{title}{Analysis and design of unconstrained nonlinear {MPC}
  schemes for finite and infinite dimensional systems}.
\newblock \bibinfo{journal}{SIAM Journal on Control and Optimization}
  \bibinfo{volume}{48}, \bibinfo{pages}{1206--1228}.
\bibitem[{Gr{\"u}ne and Pannek(2011)}]{grune2011nonlinear}
\bibinfo{author}{Gr{\"u}ne, L.}, \bibinfo{author}{Pannek, J.},
  \bibinfo{year}{2011}.
\newblock \bibinfo{title}{Nonlinear model predictive control}, in:
  \bibinfo{booktitle}{Nonlinear Model Predictive Control}.
  \bibinfo{publisher}{Springer-Verlag London Limited}, p. \bibinfo{pages}{174}.
\bibitem[{Hassan et~al.(2017)Hassan, Cipcigan and
  Jenkins}]{SANIHASSAN2017422_EMS}
\bibinfo{author}{Hassan, A.S.}, \bibinfo{author}{Cipcigan, L.},
  \bibinfo{author}{Jenkins, N.}, \bibinfo{year}{2017}.
\newblock \bibinfo{title}{Optimal battery storage operation for {PV} systems
  with tariff incentives}.
\newblock \bibinfo{journal}{Applied Energy} \bibinfo{volume}{203},
  \bibinfo{pages}{422 -- 441}.
\newblock \URLprefix
  \url{http://www.sciencedirect.com/science/article/pii/S030626191730778X},
  \DOIprefix\doi{https://doi.org/10.1016/j.apenergy.2017.06.043}.
\bibitem[{{HES Documentation}()}]{App_heat_gain}
\bibinfo{author}{{HES Documentation}}, .
\newblock \bibinfo{title}{Home energy saver \& score: Engineering
  documentation, internal gains}.
\newblock
  \bibinfo{note}{\href{http://hes-documentation.lbl.gov/calculation-methodology/calculation-of-energy-consumption/heating-and-cooling-calculation/internal-gains}
  {\texttt{http://hes-documentation.lbl.gov/cal\\culation-methodology/calculation-of-energy-consumption/heating-a\\nd-cooling-calculation/internal-gains}}
  [Accessed on 03-12-2018]}.
\bibitem[{{IESO}(2017)}]{Co2_factor}
\bibinfo{author}{{IESO}}, \bibinfo{year}{2017}.
\newblock \bibinfo{title}{Conservation {F}ramework {M}id-term {R}eview -
  {W}ebinar \#3 : {IESO} {R}esponse to {S}takeholder {Feedback}, {S}eptember 7,
  2017}.
\newblock
  \bibinfo{howpublished}{\url{http://www.ieso.ca/-/media/files/ieso/document-library/engage/cf/cf-20170921-response-to-feedback.pdf?la=en}}.
\newblock \bibinfo{note}{[Accessed on 21-12-2017]}.
\bibitem[{{IESO : Feed-in Tariff Program}()}]{FIT_Ontario}
\bibinfo{author}{{IESO : Feed-in Tariff Program}}, .
\newblock \bibinfo{title}{{FIT 5 Contract Offer Summary}}.
\newblock
  \bibinfo{howpublished}{\url{http://www.ieso.ca/-/media/Files/IESO/Document-Library/FIT/archive/FIT-5-Contract-Offers-Summary.pdf?la=en}}.
\newblock \bibinfo{note}{[Accessed on 22-09-2018]}.
\bibitem[{Kegel et~al.(2014)Kegel, Tamasauskas and
  Sunye}]{kegel2014integration}
\bibinfo{author}{Kegel, M.}, \bibinfo{author}{Tamasauskas, J.},
  \bibinfo{author}{Sunye, R.}, \bibinfo{year}{2014}.
\newblock \bibinfo{title}{{Integration and evaluation of innovative and
  renewable energy technologies in a Canadian Mid-rise Apartment}}, in:
  \bibinfo{booktitle}{9th System Simulation in Buildings Conference, paper}.
\bibitem[{Mazzeo et~al.(2018)Mazzeo, Oliveti, Baglivo and
  Congedo}]{mazzeo2018energy}
\bibinfo{author}{Mazzeo, D.}, \bibinfo{author}{Oliveti, G.},
  \bibinfo{author}{Baglivo, C.}, \bibinfo{author}{Congedo, P.M.},
  \bibinfo{year}{2018}.
\newblock \bibinfo{title}{Energy reliability-constrained method for the
  multi-objective optimization of a photovoltaic-wind hybrid system with
  battery storage}.
\newblock \bibinfo{journal}{Energy} \bibinfo{volume}{156},
  \bibinfo{pages}{688--708}.
\bibitem[{Meral and Dincer(2011)}]{meral2011review}
\bibinfo{author}{Meral, M.E.}, \bibinfo{author}{Dincer, F.},
  \bibinfo{year}{2011}.
\newblock \bibinfo{title}{A review of the factors affecting operation and
  efficiency of photovoltaic based electricity generation systems}.
\newblock \bibinfo{journal}{Renewable and Sustainable Energy Reviews}
  \bibinfo{volume}{15}, \bibinfo{pages}{2176--2184}.
\bibitem[{{Mitsubishi Electric : Cooling \& Heating}()}]{Outdoor_unit}
\bibinfo{author}{{Mitsubishi Electric : Cooling \& Heating}}, .
\newblock \bibinfo{title}{{R410a Outdoor Unit - MXZ-4C36NAHZ}}.
\newblock
  \bibinfo{howpublished}{\url{http://meus1.mylinkdrive.com/item/MXZ-4C36NAHZ.html}}.
\newblock \bibinfo{note}{[Accessed on 16-12-2018]}.
\bibitem[{{Mitsubishi Electric: Cooling \& Heating}(a)}]{Indoor_unit_Bedrooms}
\bibinfo{author}{{Mitsubishi Electric: Cooling \& Heating}}, a.
\newblock \bibinfo{title}{R410a {I}ndoor {W}all-mounted {U}nit - {MSZ-FH06NA}}.
\newblock
  \bibinfo{note}{\href{http://meus1.mylinkdrive.com/item/MSZ-FH06NA.html}
  {\texttt{http://meus1.mylinkdrive.com/item/MSZ-FH06NA.html}} [Accessed on
  16-12-2018]}.
\bibitem[{{Mitsubishi Electric: Cooling \& Heating}(b)}]{Indoor_unit_LK}
\bibinfo{author}{{Mitsubishi Electric: Cooling \& Heating}}, b.
\newblock \bibinfo{title}{{R410a Indoor Wall-mounted Unit - MSZ-FH12NA}}.
\newblock
  \bibinfo{note}{\href{http://meus1.mylinkdrive.com/item/MSZ-FH12NA.html}
  {\texttt{http://meus1.mylinkdrive.com/item/MSZ-FH12NA.html}} [Accessed on
  16-12-2018]}.
\bibitem[{{Ontario Energy Board}()}]{Eprice}
\bibinfo{author}{{Ontario Energy Board}}, .
\newblock \bibinfo{title}{Electricity rates \& prices}.
\newblock
  \bibinfo{howpublished}{\url{https://www.oeb.ca/rates-and-your-bill/electricity-rates}}.
\newblock \bibinfo{note}{[Accessed on 24-06-2018]}.
\bibitem[{Seal(2019)}]{Seal_thesis}
\bibinfo{author}{Seal, S.}, \bibinfo{year}{2019}.
\newblock \bibinfo{title}{{Model Predictive Control of Electric Building Energy
  and Heating Systems}}.
\newblock Ph.D. thesis. McGill University.
\bibitem[{Seal et~al.(2019)Seal, Dehkordi and Boulet}]{Seal_2019}
\bibinfo{author}{Seal, S.}, \bibinfo{author}{Dehkordi, V.R.},
  \bibinfo{author}{Boulet, B.}, \bibinfo{year}{2019}.
\newblock \bibinfo{title}{{Coordination of Radiant Floor and Baseboard Heating
  Systems: Sequential and Simultaneous MPC schemes}}.
\newblock \bibinfo{journal}{Science and Technology for the Built Environment}
  \bibinfo{volume}{25}, \bibinfo{pages}{1419--1436}.
\newblock \URLprefix \url{https://doi.org/10.1080/23744731.2019.1626166},
  \DOIprefix\doi{10.1080/23744731.2019.1626166},
  \href{http://arxiv.org/abs/https://doi.org/10.1080/23744731.2019.1626166}{\tt
  arXiv:https://doi.org/10.1080/23744731.2019.1626166}.
\bibitem[{{Solar Energy Laboratory, Univ. of
  Wisconsin-Madison}(2012)}]{RI_TRNSYS_ref}
\bibinfo{author}{{Solar Energy Laboratory, Univ. of Wisconsin-Madison}},
  \bibinfo{year}{2012}.
\newblock \bibinfo{title}{{TRNSYS 17 main documentation, Volume 04:
  Mathematical Reference}}.
\newblock \bibinfo{organization}{{TRNSYS 17: Thermal Energy System Specialists,
  LLC}}. \bibinfo{address}{Madison, Wisconsin}.
\bibitem[{Stelpro()}]{BB}
\bibinfo{author}{Stelpro}, .
\newblock \bibinfo{title}{{Stelpro Electric Baseboard}}.
\newblock
  \bibinfo{howpublished}{\url{https://www.stelpro.com/en-CA/electric-baseboard}}.
\newblock \bibinfo{note}{[Accessed on 16-12-2018]}.
\bibitem[{Stephenson(March 1967)}]{Qsol_heat_gain_factor}
\bibinfo{author}{Stephenson, D.G.}, \bibinfo{year}{March 1967}.
\newblock \bibinfo{title}{{Tables of Solar Altitude, Azimuth, Intensity and
  Heat Gain Factors for Latitudes From 43 to 55 Degrees North}}.
\newblock \bibinfo{type}{{Technical Report}}. National Research Council of
  Canada. Division of Building Research.
\newblock \URLprefix \url{https://doi.org/10.4224/20378809}.
\bibitem[{{Tesla Powerwall}()}]{HB_powerwall}
\bibinfo{author}{{Tesla Powerwall}}, .
\newblock \bibinfo{title}{{Powerwall performance and mechanical
  specifications}}.
\newblock
  \bibinfo{howpublished}{\url{https://www.tesla.com/sites/default/files/pdfs/powerwall/Powerwall\%202_AC_Datasheet_en_northamerica.pdf}}.
\newblock \bibinfo{note}{[Accessed on 21-04-2019]}.
\bibitem[{{TESSLibs 17: Component Libraries for the TRNSYS Simulation
  Environment}(2013)}]{PV_TRNSYS_ref}
\bibinfo{author}{{TESSLibs 17: Component Libraries for the TRNSYS Simulation
  Environment}}, \bibinfo{year}{2013}.
\newblock \bibinfo{title}{{Volume 03: Electrical Library Mathematical
  Reference}}.
\newblock \bibinfo{organization}{{TESS - Thermal Energy Systems Specialists}}.
  \bibinfo{address}{Madison, Wisconsin}.

\end{thebibliography}


%
%

\end{document}